\documentclass[preprint,3p,times,preprintnumber]{elsarticle}
\pdfoutput = 1
\usepackage{bbm}
\usepackage{amsmath}
\usepackage{amssymb}
\usepackage{amsfonts}
\usepackage{mathrsfs}
\usepackage{graphicx}
\usepackage[11pt]{moresize}
\usepackage[normalem]{ulem}
\usepackage[dvipsnames]{xcolor}
\usepackage[hidelinks]{hyperref}
\usepackage{orcidlink}
\usepackage{relsize}
\usepackage{lineno}
\usepackage{appendix}

\hypersetup
{
 bookmarks  = true,      % show bookmarks bar?
 unicode    = false,     % non-Latin characters in Acrobat?s bookmarks
 colorlinks = true,      % false: boxed links; true: colored links
 linkcolor  = red,       % color of internal links
 citecolor  = blue,      % color of links to bibliography
 filecolor  = black,     % color of file links
 urlcolor   = blue,      % color of external links
}

\newcommand{\amt}{{\text a}_{\mu\tau}}
\newcommand{\emt}{\varepsilon_{\mu\tau}}

\journal{Physics Letter B}

\begin{document}

\begin{flushright}
IP/BBSR/2022-3, TIFR/TH/22-19
\end{flushright}

\title{Discriminating between Lorentz violation and non-standard interactions using core-passing atmospheric neutrinos at INO-ICAL}

\address[s1]{Institute of Physics, Sachivalaya Marg, Sainik School Post, Bhubaneswar 751005, India}
\address[s2]{Homi Bhabha National Institute, Anushakti Nagar, Mumbai 400085, India}
\address[s3]{Applied Nuclear Physics Division, Saha Institute of Nuclear Physics, Bidhannagar, Kolkata 700064, India}
\address[s4]{Department of Physics \& Wisconsin IceCube Particle Astrophysics Center, University of Wisconsin, Madison, WI 53706, USA}
\address[s5]{Tata Institute of Fundamental Research, Homi Bhabha Road, Colaba, Mumbai 400005, India}

\author[s1,s2]{\orcidlink{0000-0001-6719-7723}Sadashiv Sahoo}
\ead{sadashiv.sahoo@iopb.res.in}
%%%%
\author[s1,s2,s3]{\orcidlink{0000-0002-8367-8401}Anil Kumar}
\ead{anil.k@iopb.res.in}
%%%%
\author[s1,s2,s4]{\orcidlink{0000-0002-9714-8866}Sanjib Kumar Agarwalla}
\ead{sanjib@iopb.res.in}
%%%%
\author[s5]{\orcidlink{0000-0001-6639-0951}Amol Dighe}
\ead{amol@theory.tifr.res.in}

%%%%
\date{\today}

\begin{abstract}
Precision measurements of neutrino oscillation parameters have provided a tremendous boost to the search for sub-leading effects due to several beyond the Standard Model scenarios in neutrino oscillation experiments. Among these, two of the well-studied scenarios are Lorentz violation (LV) and non-standard interactions (NSI), both of which can affect neutrino oscillations significantly. We point out that, at a long-baseline experiment where the neutrino oscillation probabilities can be well-approximated by using the line-averaged constant matter density, the effects of these two scenarios can mimic each other. This would allow the limits obtained at such an experiment on one of the above scenarios to be directly translated to the limits on the other scenario. However, for the same reason, it would be difficult to distinguish between LV and NSI at a long-baseline experiment. We show that the observations of atmospheric neutrinos, which travel a wide range of baselines and may encounter sharp density changes at the core-mantle boundary, can break this degeneracy. We observe that identifying neutrinos and antineutrinos separately, as can be done at INO-ICAL, can enhance the capability of atmospheric neutrino experiments to discriminate between these two new-physics scenarios.

\end{abstract}


\maketitle

%=============================%
\section{Introduction}
%=============================%
The Standard Model (SM) of particle physics is the most successful theory of elementary particles and their fundamental interactions~\cite{Zyla:2020zbs}. It can explain most of the particle properties and associated phenomena with a high accuracy. However, there are certain experimental observations that cannot be accommodated in the SM; one of the most established ones is the phenomenon of neutrino oscillations. Over the last two decades, several pioneering experiments involving solar~\cite{Super-Kamiokande:1998oic, Super-Kamiokande:2001bfk, SNO:2001kpb, Super-Kamiokande:2002ujc,Super-Kamiokande:2005wtt, Super-Kamiokande:2008ecj, Super-Kamiokande:2010tar, SNO:2011hxd}, atmospheric~\cite{Achar:1965ova, Fukuda:1998mi, Ashie2004, IceCube:2014flw, Super-Kamiokande:2017yvm}, reactor~\cite{Eguchi:2002dm, Araki:2004mb, An:2012eh, Ahn:2012nd, KamLAND:2013rgu, RENO:2018dro, DayaBay:2018yms, DoubleChooz:2019qbj}, and accelerator~\cite{K2K:2004iot, Adamson:2008zt, MINOS:2013xrl, MINOS:2013utc, T2K:2019bcf, NOvA:2019cyt} neutrinos have established that neutrinos change their flavor during propagation. This demands that neutrinos have mass and they mix with each other.
 
To accommodate the observed tiny neutrino masses and large mixing angles, we need physics beyond the Standard Model (BSM). The study of neutrino oscillations can reveal the nature of the BSM physics which affects production, propagation, and detection of neutrinos in oscillation experiments. Next generation experiments will measure the mass-mixing parameters with the precision of a few per cent. This will allow us to probe multiple BSM scenarios, whose effects on neutrino oscillations may be small~\cite{Arguelles:2019xgp}.

Two such BSM scenarios widely discussed in the literature are (i) neutrino interactions involving Lorentz violation (LV)~\cite{Colladay:1996iz, Auerbach:2005tq,  Adamson:2008aa, Adamson:2010rn, Abbasi:2010kx, AguilarArevalo:2011yi, Kostelecky:2008ts, Adamson:2012hp, Adamson:2008aa, Abe:2012gw, Rebel:2013vc, Diaz:2013iba, Diaz:2016fqd, Abe:2014wla, Abe:2017eot, Adey:2018qsd, Aartsen:2017ibm, Sahoo:2021dit} and (ii) non-standard interactions (NSI) of neutrinos with ambient matter~\cite{Wolfenstein:1977nu, Ohlsson:2012kf,  Fornengo:2001pm, Kopp:2010qt, Super-Kamiokande:2011dam, Choubey:2015xha, Salvado:2016uqu, IceCube:2017zcu, Farzan:2017xzy, Khatun:2019tad, Kumar:2021lrn, HernandezRey:2021qac, IceCube:2021euf, ANTARES:2021crm, Denton:2021rgt}. 
Note that the origins of new interactions of neutrino in these two scenarios are completely different: LV arises from the interactions of neutrinos with the spacetime itself, while NSI effects emerge from neutrino interactions with matter. The LV can manifest itself in vacuum as well as in matter, whereas the neutral-current NSI effects appear only during neutrino propagation in matter. Here, we investigate the imprints of both these BSM scenarios on neutrino oscillations.

In spite of the intrinsic differences between them, both the above scenarios affect the neutrino propagation inside Earth in a very similar manner. So much so that, in a long-baseline neutrino oscillation experiment, the effective Hamiltonians of these two scenarios can almost exactly mimic each other. Therefore, if one of these two scenarios is realized in Nature, it would be difficult to rule out the other from observations at these experiments. 

In this work, we demonstrate for the first time that the degeneracy between these two scenarios can be broken by atmospheric neutrino experiments having access to a wide range of baselines inside Earth. The observations of neutrinos and antineutrinos passing through the core of Earth would play a critical role in discriminating between these two BSM scenarios. We further find that the sensitivity towards distinguishing between them would be enhanced if neutrinos and antineutrinos can be detected separately. This would be possible with the charge identification (CID) capability of a detector like the proposed iron calorimeter (ICAL) at the India-based Neutrino Observatory (INO)~\cite{ICAL:2015stm}. Our aim in this paper is to determine the sensitivity of the ICAL-INO detector to discriminate between LV and NSI using the state-of-the-art simulation tools developed for ICAL that exploit its excellent energy and angular resolutions in the multi-GeV energy range, and its CID capability.  

%=============================%
\section{Lorentz Violation (LV)}
%=============================%
The Lorentz symmetry is a fundamental ingredient of the SM, and indeed, of local quantum field theories in general. However, there are a few proposed models in string theory~\cite{Polyakov:1987ez, Kostelecky:1988zi, Kostelecky:1989jp, Kostelecky:1990pe, Kostelecky:1991ak, Kostelecky:1995qk, Kostelecky:1999mu, Kostelecky:2000hz} and loop quantum gravity~\cite{Gambini:1998it, Alfaro:2002xz, Sudarsky:2002ue, Amelino-Camelia:2002aqz, Ng:2003jk} which give rise to LV. We shall focus on a scenario in which the Lorentz Symmetry is broken spontaneously, giving nonzero vacuum expectation value (vev) to a 4-vector ${\text a}^\lambda$. Here $\lambda$ is the spacetime index. The couplings of ${\text a}^\lambda$ with neutrinos are flavor-dependent, and hence the Lagrangian density of the LV interaction may be written as~\cite{Kostelecky:2011gq, Barenboim:2018ctx, Sahoo:2021dit}
\begin{align}
\mathcal{L}_{\rm LV} &= -{\text a}^{\mathsmaller{\lambda}}_{\alpha\beta}
(\overline{\nu}_\alpha\gamma_{\mathsmaller{\lambda}}P_L \nu_\beta)\,,
\label{Eq:L-LV}
\end{align}
where ${\text a}^\lambda_{\alpha\beta}$ combines the information on the vev and couplings of ${\text a}^\lambda$. Here $\alpha,\beta$ are the flavor indices and the operator $P_L$ corresponds to the left chiral projection. The hermiticity of interactions imposes ${\text a}^\lambda_{\beta\alpha} = ({\text a}^\lambda_{\alpha\beta})^*$. Note that the above interaction also breaks the Charge conjugation - Parity - Time reversal (CPT) symmetry, since the elements of ${\text a}^\lambda$ change sign under CPT transformation~\cite{Kostelecky:2003cr, Sahoo:2021dit}. Our scenario may therefore also be termed as a CPT-violating scenario, which guarantees LV automatically~\cite{Greenberg:2002uu}. Since there are strong constraints on the observed LV, it is expected that the elements of ${\text a}^\lambda$ are suppressed by the Planck scale $M_P$~\cite{Appelquist:1974tg, Colladay:1996iz, Colladay:1998fq, Kostelecky:2000mm, Kostelecky:2003cr, Kostelecky:2003fs, Bluhm:2005uj}. Note that for antineutrinos, ${\text a}^\lambda_{\alpha\beta}\to -({\text a}^\lambda_{\alpha\beta})^*$.

We work in an approximately inertial frame and consider only the timelike component of ${\rm a}^\lambda$ to be nonzero; i.e. ${\rm a}^0 \ne 0$. The Sun-centered celestial-equatorial (SCCE) frame~\cite{Kostelecky:2008ts} can be taken to be such a frame when the small effects due to gravity and boost due to the Earth's motion are ignored. In this frame, the total effective Hamiltonian of ultra-relativistic left-handed neutrinos passing through Earth can be written in the 3$\nu$ flavor basis as
\begin{align}
\mathcal{H}_{\rm LV} &= \frac{1}{2E} \mathbbm{U} \mathbbm{M}^2 \mathbbm{U}^\dagger
+ \sqrt{2}G_{F}N_{e}\widetilde{\mathbb{I}} +
\mathbbm{A} \;,
\label{Eq:1.2}
\end{align}
where $\mathbbm{U}$ is the neutrino mixing matrix, also called as the Pontecorvo - Maki - Nakagawa - Sakata (PMNS) matrix, while $\mathbbm{M}^2$ is the diagonal matrix with elements $(0,\,\Delta m_{21}^2,\,\Delta m_{31}^2)$. The first term represents the Hamiltonian in vacuum in the absence of any LV interaction. The second term incorporates the effective matter potential experienced by neutrinos as they propagate through matter with electron density $N_e$, due to their charged-current interactions with ambient electrons. Here $G_F$ is the Fermi constant and $\tilde{\mathbb{I}}$ is the diagonal matrix with elements $(1,\,0,\,0)$. In terms of the density $\rho$ of the medium through which the neutrinos propagate, one may write
\begin{align}
\sqrt{2}G_F N_e \approx 7.6 \times 10^{-23}\cdot Y_e\cdot\rho\left({\rm g/cm^3}\right)
\mbox{ GeV} \; ,
\end{align}
where $Y_e$ is the electron-number fraction in the medium. The last term arises from Eq.~(\ref{Eq:L-LV}), with $\mathbbm{A}$  being the LV matrix in the neutrino flavor space with its elements $ {\text a}^0_{\alpha \beta}$. (Henceforth, we shall omit the superscript `0' for the sake of brevity.) For antineutrino, $\mathbbm{U} \to \mathbbm{U}^*$, while both the second and third terms change sign. Note that, while the third term is intrinsically CPT-violating, the second term gives rise to matter-induced  CPT violation~\cite{Jacobson:2003wc, Ohlsson:2014cha}.

%=============================%
\begin{figure}[t]
\centering
\includegraphics[width=0.45\textwidth]{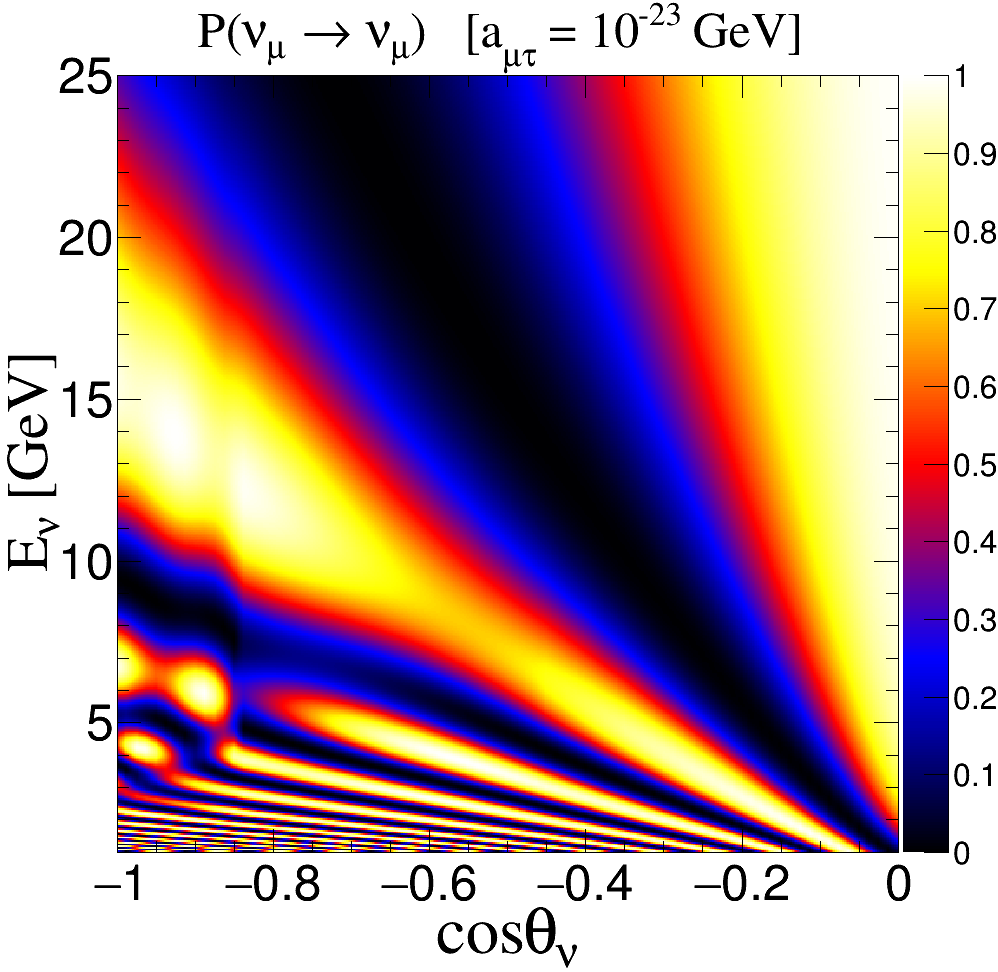}
\hspace{0.2 cm}
\includegraphics[width=0.45\textwidth]{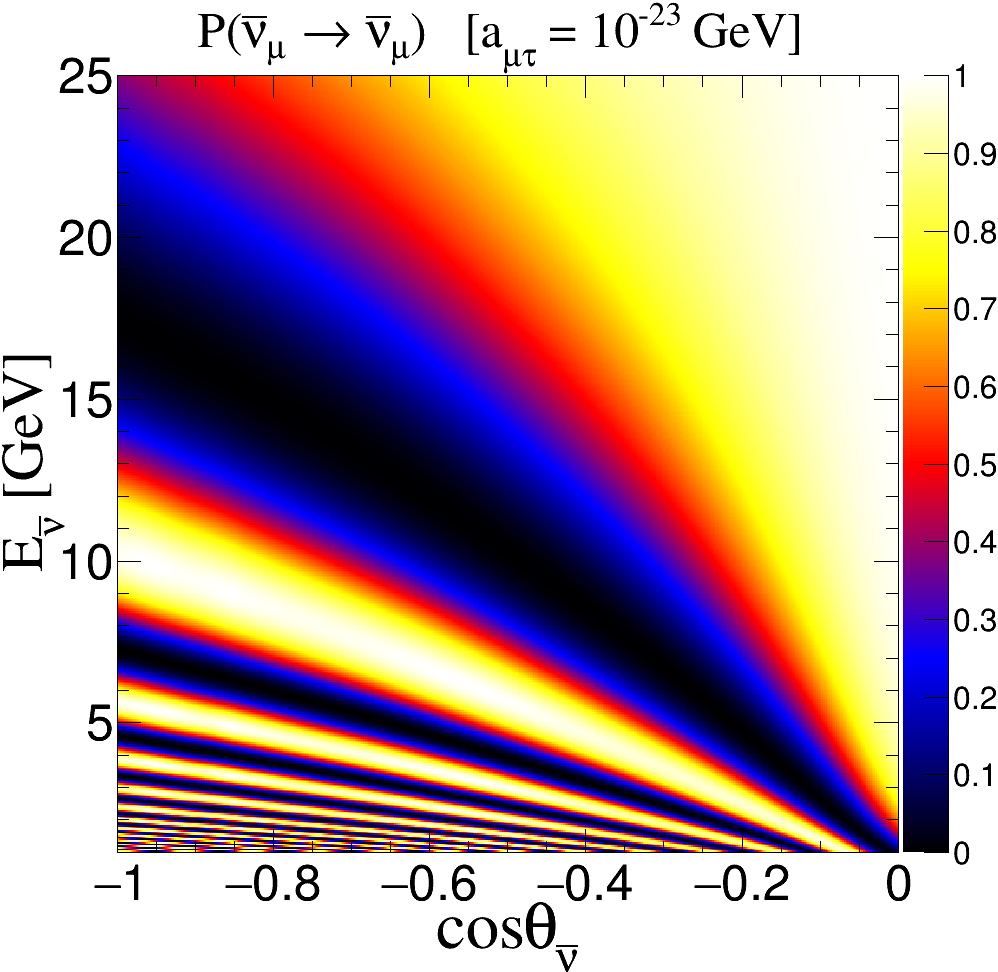}
\caption{Survival probabilities in $\big(\cos\theta_\nu, E_\nu\big)$ plane, for $\amt = 10^{-23}$ GeV. Left (Right) panel is for upward-going $\nu_\mu$ ($\bar\nu_\mu$).}
\label{fig:oscillogram-amunu}
\end{figure}
%=============================%

We focus on charged-current muon events at an atmospheric neutrino experiment that is sensitive to multi-GeV neutrinos. More than 98\% of such events arise via $\nu_\mu$ disappearance. The dominant LV corrections to the relevant survival probability $P(\nu_\mu \to \nu_\mu)$ stem from $\amt$~\cite{Kopp:2007ne, Sahoo:2021dit}. For real $\amt$, the current experimental constraint from Super-K is $|\amt| \le 0.65 \times 10^{-23}$ GeV at 95\% C.L.~\cite{Abe:2014wla} and from IceCube is $|\amt| \le 0.29 \times 10^{-23}$ GeV at 99\% C.L.~\cite{Aartsen:2017ibm}. In Fig.~\ref{fig:oscillogram-amunu}, we show the effect of a benchmark value of $\amt = 10^{-23}$ GeV on the survival probability of upward-going multi-GeV $\nu_\mu$ and $\bar\nu_\mu$, in the plane of zenith angle and energy $\left(\cos\theta_\nu,\,E_\nu \right)$. The oscillation valley, i.e., the central black region with the smallest survival probability, would be an almost triangular strip in the absence of any BSM physics~\cite{Kumar:2021lrn, Kumar:2020wgz}. It is observed that the LV effects bend this valley in opposite directions for $\nu_\mu$ and $\bar\nu_\mu$. The strong matter effects at $\cos\theta_\nu < -0.85$, especially at low energies, arise from the $\sqrt{2} G_F N_e \widetilde{\mathbbm{I}}$ term.

For all the numerical results in this work, we use the benchmark oscillation parameters as $\sin^2 2\theta_{12} = 0.855$, $\sin^2 2\theta_{13} = 0.0875$, $\sin^2 \theta_{23} = 0.5$, $\Delta m^2_{32} = + 2.46 \times 10^{-3}$ eV$^2$ (normal mass ordering), $\Delta m^2_{21} = 7.4 \times 10^{-5}$ eV$^2$, and $\delta_{\rm CP} = 0$~\cite{NuFIT,Esteban:2020cvm, deSalas:2020pgw}. We use the Preliminary Reference Earth Model (PREM)~\cite{Dziewonski:1981xy} for the Earth density profile.

%=============================%
\section{Non-Standard Interactions (NSI)}
%=============================%
In this scenario, neutrinos undergo additional coherent forward scattering with the matter fermions (up-quark $u$, down-quark $d$, electron $e$) due to new interactions. These NSIs may originate from new physics at an energy scale $\Lambda$ higher than the scale of electroweak interactions $\big(m_W,\, m_Z\big)$. They may be written in the effective field theory language in terms of four-fermion operators with mass-dimension six~\cite{Weinberg:1979sa}.
%=============================%
\begin{figure}[t]
\centering
\includegraphics[width=0.45\textwidth]{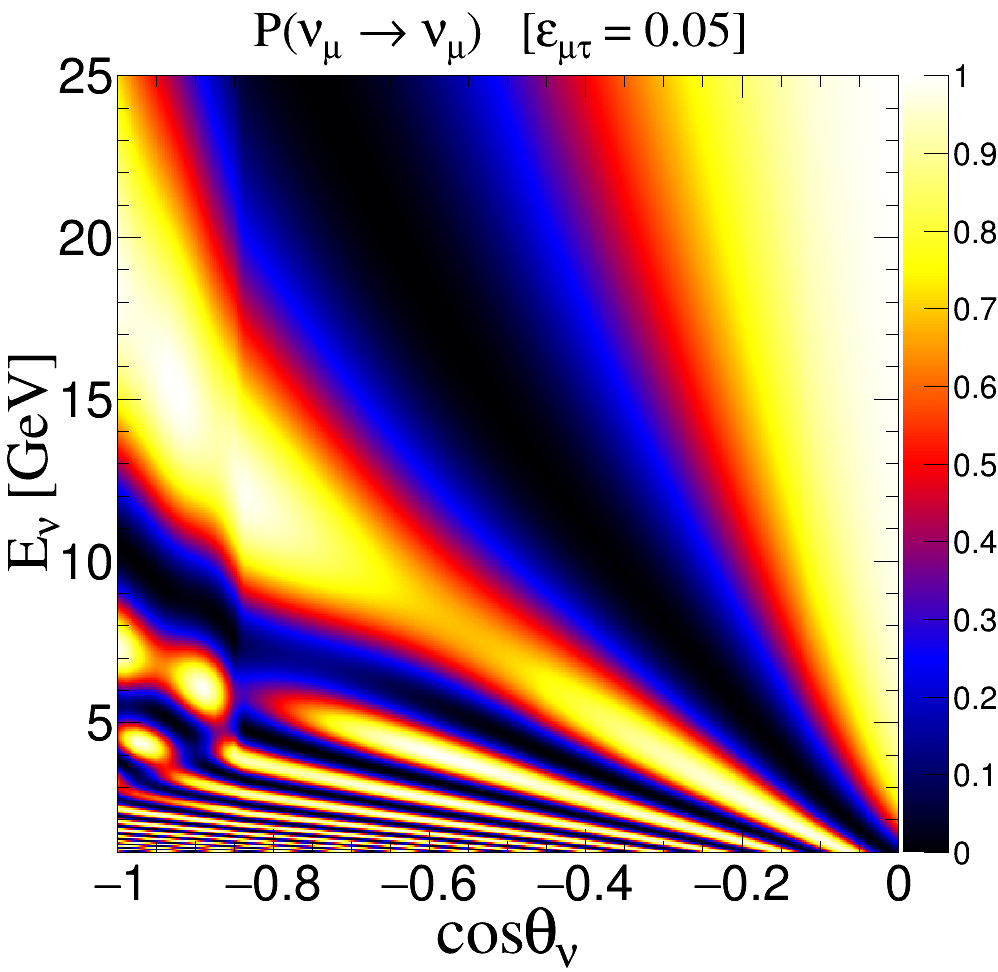}
\hspace{0.2 cm}
\includegraphics[width=0.45\textwidth]{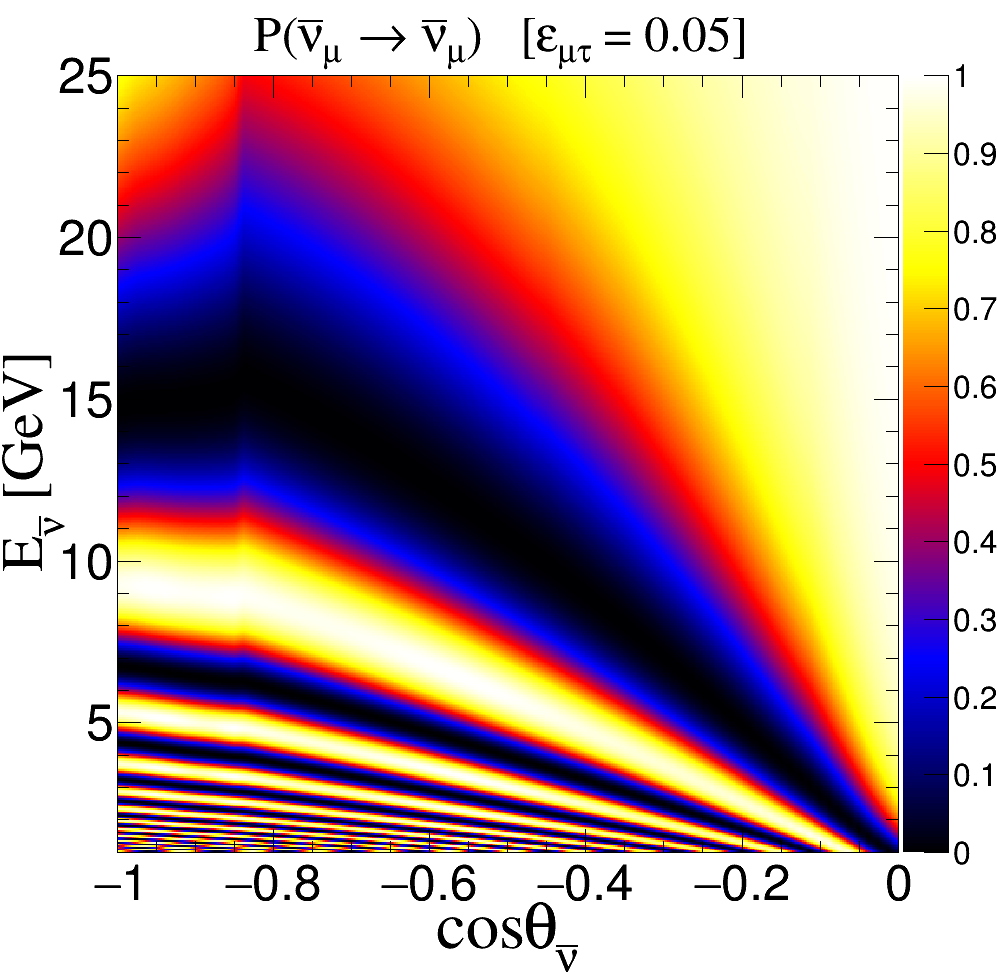}
\caption{Survival probabilities in $\big(\cos\theta_\nu, E_\nu\big)$ plane, for $\emt = 0.05$. Left (Right) panel is for upward-going $\nu_\mu$ ($\bar\nu_\mu$).}
\label{fig:oscillogram-epsmunu}
\end{figure}
%=============================%
In this study, we focus on the neutral-current NSI, for which the Lagrangian density is written as 
\begin{align}
\mathcal{L}_{\rm NSI} &= -2\sqrt{2} \, G_F \,
\varepsilon^{f\, C}_{\alpha\beta}
\big(\bar\nu_\alpha \gamma^{\mathsmaller{\eta}} P_{\mathsmaller{L}}\nu_\beta\big)
\big(\bar f \gamma_{\mathsmaller{\eta}}  P_C  f\big),
\label{Eq:2.1}
\end{align}
where $f$ are the matter fermions, $C \in \{L, R\}$, and $P_C$ are the corresponding chiral projections operators. The hermiticity of the interaction imposes $\varepsilon^{f\, C}_{\beta\alpha} = (\varepsilon^{f\, C}_{\alpha\beta})^*$. The strength of interaction is expected to be supressed by $(m_W/\Lambda)^2$, which is reflected in the smallness of $\varepsilon_{\alpha\beta}$. For the Earth matter, we assume $N_p \approx N_n = N_e$, which leads to $N_u \approx N_d \approx 3N_e$. Therefore, the effective NSI parameter $\varepsilon_{\alpha\beta} \approx \varepsilon^e_{\alpha\beta} + 3\varepsilon^u_{\alpha\beta} + 3\varepsilon^d_{\alpha\beta}$, where $\varepsilon_{\alpha\beta}^f \equiv \varepsilon_{\alpha\beta}^{f\,L} + \varepsilon_{\alpha\beta}^{f\,R}$. The effective Hamiltonian for neutrinos propagating through Earth can be expressed in the 3$\nu$ flavor basis as
\begin{align}
\mathcal{H}_{\rm NSI} & = \frac{1}{2E} \mathbbm{U} \mathbbm{M}^2 \mathbbm{U}^\dagger
 + \sqrt{2}G_{F}N_{e}\widetilde{\mathbb{I}} + \sqrt{2}G_{F}N_{e}\, \mathcal{E}\, ,
\label{Eq:2.2}
\end{align}
where $\mathcal{E}$ is a matrix with elements $\varepsilon_{\alpha\beta}$. For antineutrino, the signs of the second and third term are reversed. The element $\emt$ is expected to affect the muon survival probabilities at the leading order~\cite{Kopp:2007ne}. The value of $|\emt|$ has been constrained from the measurements at Super-K~\cite{Super-Kamiokande:2011dam}, IceCube-DeepCore~\cite{IceCube:2017zcu,IceCube:2021euf}, and ANTARES~\cite{ANTARES:2021crm}. Recently, the most stringent constraint has been obtained using the TeV-Scale $\nu_\mu$ disappearance data from the IceCube experiment that corresponds to $|\emt| \lesssim 0.01$ at 90\% C.L.~\cite{IceCube:2022ubv}.

In Fig.~\ref{fig:oscillogram-epsmunu}, we show the effect of non-zero $\emt$ at a representative value $\emt = +0.05$. One may observe that for $\cos\theta_\nu \gtrsim -0.85$, the bending of the oscillation valley is quite similar to that due to LV. However, for upward-going neutrinos that have passed through the core of Earth, i.e. for $\cos\theta_\nu \lesssim -0.85$, the figures~\ref{fig:oscillogram-amunu} and~\ref{fig:oscillogram-epsmunu} show major differences. The above similarities manifest the degeneracy between the LV and NSI scenarios, while the differences provide the key to its resolution.

%=============================%
\section{The LV-NSI degeneracy}
%=============================%
The comparison between Eqs.~(\ref{Eq:1.2}) and (\ref{Eq:2.2}) shows that, if
\begin{align}
\mathbbm{A} = \sqrt{2}\, G_F\, N_e\, \mathcal{E} \; ,
\label{eq:degen}
\end{align}
the effective Hamiltonians of the LV and NSI scenarios are identical~\cite{Diaz:2015dxa, KumarAgarwalla:2019gdj, Sahoo:2021dit}. In such a case, the effects of the LV scenario will be completely mimicked by NSI with appropriate parameter values, and vice versa. This would imply that, if the data prefer a particular value of $\varepsilon_{\mu\tau}$ in the NSI scenario, it would also automatically prefer the corresponding value of $\amt$ in the LV scenario, and vice versa. 

\begin{figure}
	\centering
	\includegraphics[width=0.5\linewidth]{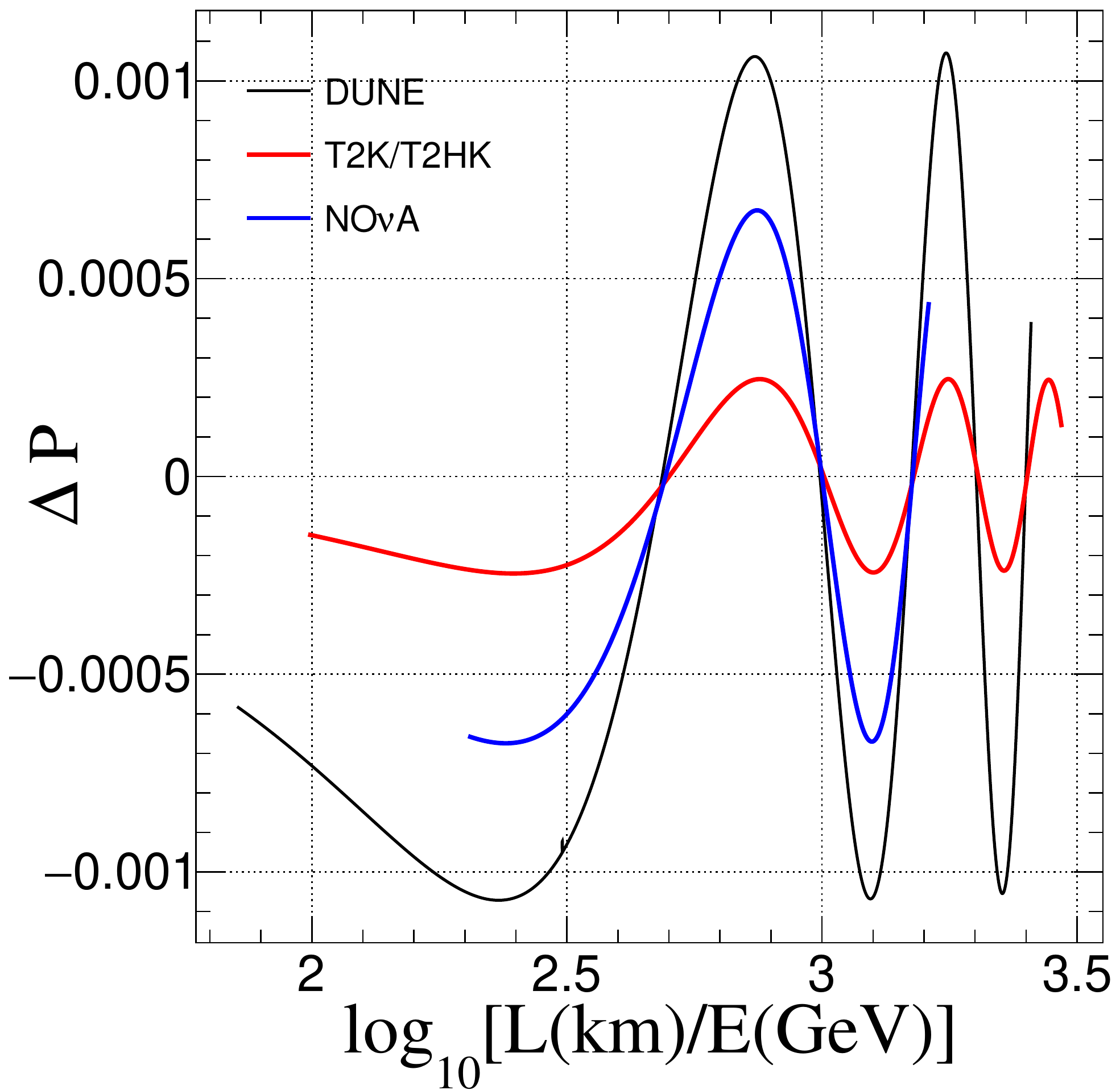}
	\caption{The difference $\Delta P$ (see Eq.~(\ref{eq:deltaP})) between the scenarios with LV ($\amt = 10^{-23}$ GeV) and NSI ($\varepsilon_{\mu\tau} = 0.092$), as a function of $L/E$. The curves are drawn for three different long-baseline experiments:  DUNE (1300 km), T2K/T2HK (295 km), and NO$\nu$A (810 km) with their corresponding neutrino energy ranges.}
	\label{fig:deltaP_lbye}
\end{figure}

When neutrinos propagate through the Earth matter, the left-hand side of Eq.~(\ref{eq:degen}) does not change, while its right-hand side changes along its path due to the $N_e$ dependence. However, at current and planned long-baseline experiments like K2K, MINOS, OPERA, T2K, NO$\nu$A, T2HK, and DUNE, the density variation encountered by neutrinos along their path is quite small, and  the oscillation probabilities can be approximated to a great accuracy by using the line-averaged constant matter density along the neutrino path~\cite{Gandhi:2004bj, Agarwalla:2008jin}. Therefore, given any value of $\mathbbm{A}$, Eq.~(\ref{eq:degen}) is always satisfied to a high accuracy for some value of $\mathcal{E}$ in these long-baseline experiments. This can be numerically confirmed by calculating the quantity 
\begin{align}
\Delta P = P\left(\mathsmaller{{\rm SM + LV}}\right) - P\left(\mathsmaller{{\rm SM + NSI}}\right),
\label{eq:deltaP}
\end{align}
for the long-baseline experiments DUNE (1300 km), T2K/T2HK (295 km), and NO$\nu$A (810 km), using complete three-neutrino oscillation probabilities in the presence of matter with the Preliminary Reference Earth Model (PREM) profile. In Fig.~\ref{fig:deltaP_lbye}, we show the quantity $\Delta P$ as a function of $L/E$, where $P \left(\rm SM+LV\right)$ has been calculated using ($\amt = 10^{-23}$ GeV,  $\varepsilon_{\mu\tau} = 0$) and $P\left(\rm SM+NSI\right)$ using ($\amt = 0$,  $\varepsilon_{\mu\tau} = 0.092$). Note that we have taken $\amt = 10^{-23}$ GeV for illustration, though the limits on this quantity are more stringent. The value of $\varepsilon_{\mu\tau} = 0.092$ has been chosen so that $\amt \approx  \sqrt{2} G_F N_e \varepsilon_{\mu\tau}$ where $N_e$ corresponds to the line-averaged density at DUNE, i.e., $\rho^{\rm DUNE}_{avg} \approx 2.85\, {\rm g/cm^3}$~\cite{Roe:2017zdw, DUNE:2021cuw}.

The figure shows that $|\Delta P| < 0.0012$ for all energies at all the major current and proposed long-baseline experiments. Since the current limits on $\amt$ and $\varepsilon_{\mu\tau}$ are smaller than the values considered in the figure, the actual value of $|\Delta P|$ would be even lower. The long-baseline experiments under consideration are not expected to reach this precision any time in near future. Therefore, the LV and NSI scenarios are indistinguishable at these experiments.

This observation has two consequences: (i) If a particular limit on $\varepsilon_{\mu\tau}$ is obtained for the NSI scenario using the data from the long-baseline experiments, the corresponding limit on $\amt$ in the LV scenario can be inferred  simply by scaling the limit on $\varepsilon_{\mu\tau}$ according to Eq.~(\ref{eq:degen}), and vice versa. (ii) If a positive signal for any one of the above scenarios is obtained at a long-baseline experiment, it would be impossible to identify whether the actual BSM scenario is LV or NSI. Thus, the degeneracy between these two scenarios is inevitable in the long-baseline setups.

%=============================%
\section{Resolving the degeneracy using atmospheric neutrinos}
%=============================%
Atmospheric neutrino experiments detect neutrinos coming from all directions. The range of distances traveled by these neutrinos through Earth is all the way from zero to $\sim 12750$ km. The line-averaged constant density (LACD) approximation is not so accurate for the neutrinos that travel large distances through the mantle. These neutrinos may also undergo the Mikheyev-Smirnov-Wolfenstein (MSW) resonance~\cite{Wolfenstein:1977nu, Mikheev:1986gs, Mikheev:1986wj}, which contributes to the deviation from LACD approximation. Moreover, the neutrinos that travel through the core encounter sharp density changes by a factor of almost 2 at the core-mantle boundary, and may be affected by the Neutrino Oscillation Length Resonance (NOLR)~\cite{Petcov:1998su,Chizhov:1998ug,Petcov:1998sg,Chizhov:1999az,Chizhov:1999he} or the parametric resonance~\cite{Akhmedov:1998ui,Akhmedov:1998xq}. As a result, the LACD approximation is badly broken for them. Thus, the condition in Eq.~(\ref{eq:degen}) is not satisfied for all neutrinos, and hence the almost-exact degeneracy between LV and NSI scenarios ceases to exist for long baselines. In addition, even when the LACD approximation is valid, since atmospheric neutrinos have multiple baselines with widely different densities, the value of $\emt$ that would mimic a given value of $\amt$ would vary from baseline to baseline. This factor would also contribute to the power of atmospheric neutrinos for distinguishing LV from NSI.

%=============================%
\begin{figure}[t]
\centering
\includegraphics[width=0.475\textwidth]{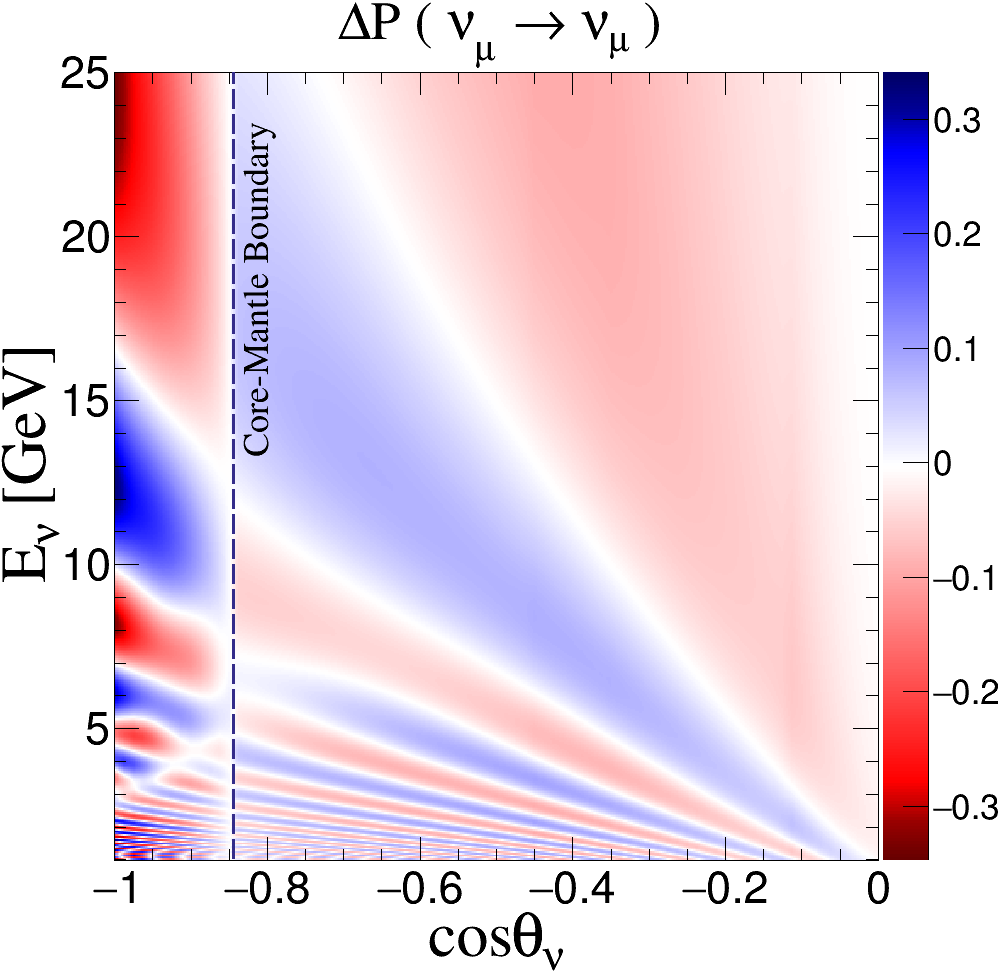}
\hspace{0.2 cm}
\includegraphics[width=0.475\textwidth]{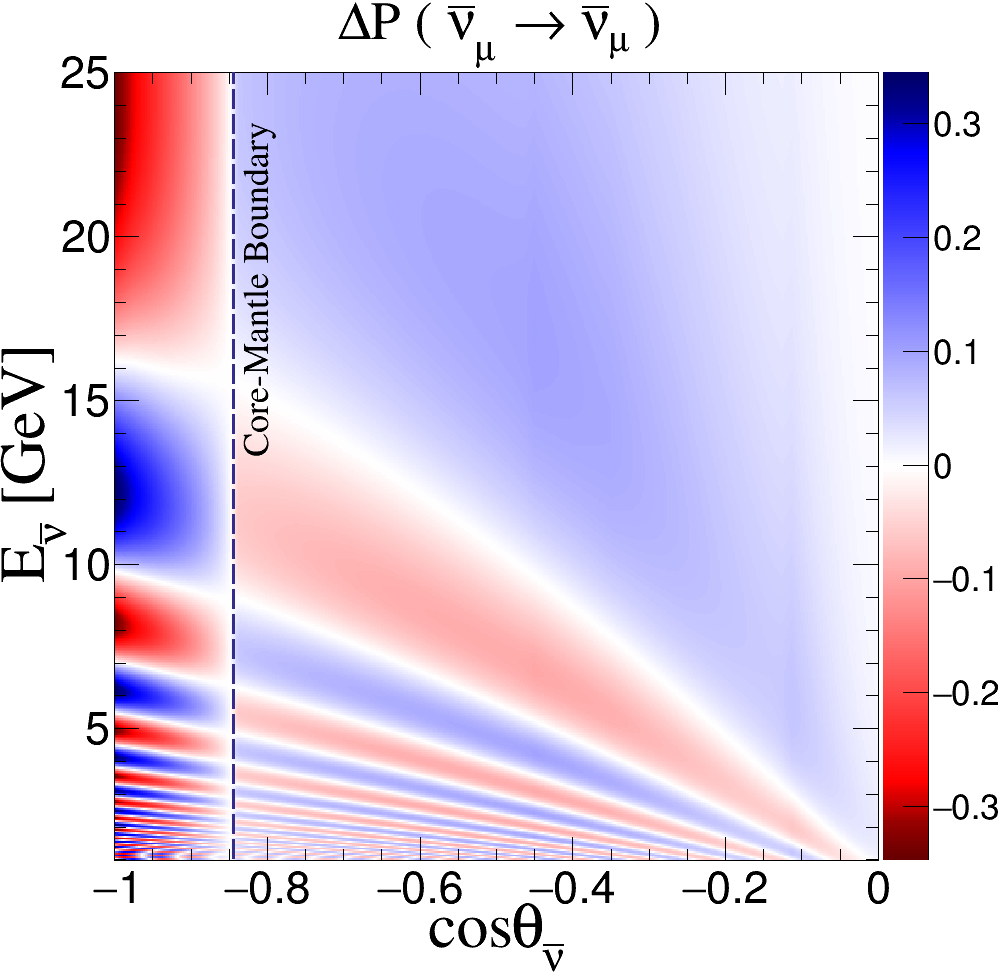}
\caption{The difference $\Delta P$ between the scenarios with LV ($\amt = 10^{-23}$ GeV) and NSI ($\varepsilon_{\mu\tau} = 0.0475$), in the $\big(\cos\theta_\nu, E_\nu\big)$ plane. The left and right panels correspond to $\nu_\mu$ and $\bar\nu_\mu$, respectively.}
\label{fig:3}
\end{figure}
%=============================%

To illustrate this point, we first consider Earth as a uniform solid sphere of average mass density $\rho^{Earth}_{avg} \approx 5.5\, {\rm g/cm^3}$, which would give the corresponding value $\emt = 0.0475$ for $\amt = 10^{-23}$ GeV, using Eq.~(\ref{eq:degen}). In Fig.~\ref{fig:3}, we show the difference $\Delta P$ between the probabilities predicted by these two scenarios, where we use three-flavor neutrino oscillation in the presence of matter effect considering the PREM profile of Earth. It is observed that for neutrinos that travel only through the mantle, i.e. for $\cos \theta_\nu \gtrsim -0.85$, we get $|\Delta P| \lesssim 0.1$. However, this difference grows for neutrinos passing through the core, reaching values as high as $|\Delta P| \approx 0.34$. Due to the significant $|\Delta P|$, the atmospheric neutrino data would be able to distinguish between the LV and NSI scenarios.

%=============================%
\section{LV vs. NSI discrimination with INO-ICAL}
%=============================%

%=============================%
\begin{figure}[t]
	\centering
	\includegraphics[width=0.6\linewidth]{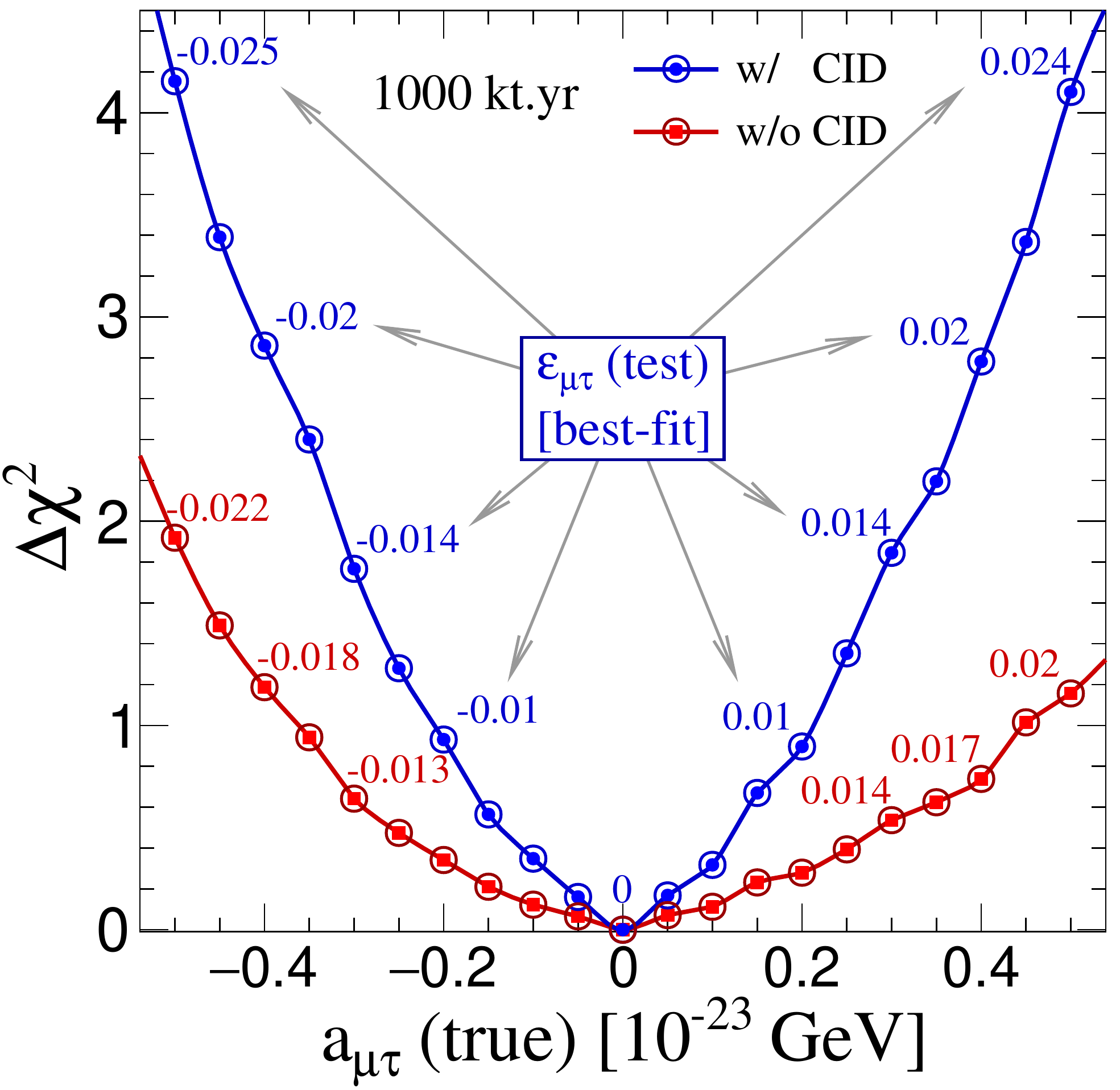}
	\caption{The minimum $\Delta \chi^2$ with the NSI hypothesis, when the actual scenario is LV with a given $\amt^{\rm true}$. The red (blue) colors indicate results without (with) CID. Some of the best-fit values of  $\emt^{\rm test}$, obtained for a given value of $\amt^{\rm true}$, are shown with numbers.}
	\label{fig:chisq}
\end{figure}
%=============================%

%=============================%
\begin{figure}[t]
	\centering
	\includegraphics[width=0.6\linewidth]{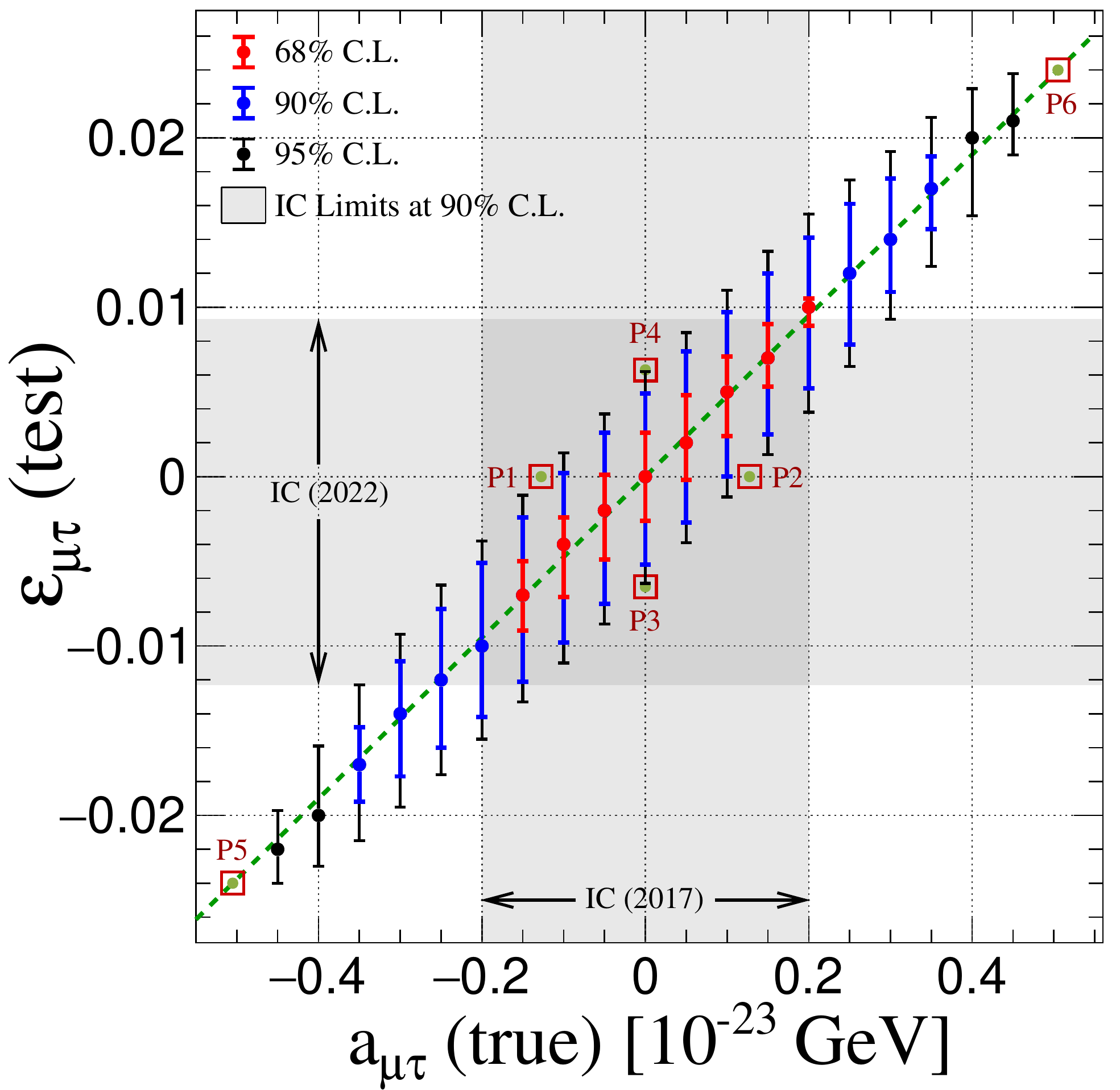}
	\caption{The values of $\Delta\chi^2$ as defined in eq. (\ref{Eq:4}), indicating how well a value of $\emt^{\rm test}$ can mimic an actual value of $\amt^{\rm true}$. The dots (red, blue, and black) are the best-fit values of $\emt^{\rm test}$ for the corresponding $\amt^{\rm true}$. The vertical red, blue, and black error-bars correspond to the regions beyond which $\emt^{\rm test}$ values may be excluded with confidence levels of 68\%, 90\%, and 95\%, respectively, with 1 degree of freedom. The green dashed line represents the $\emt^{\rm test}$ values that can mimic $\amt^{\rm true}$ with the approximation of a uniform Earth density (see eq. (\ref{eq:degen})). The vertical (horizontal) gray band corresponds to the current 90\% C.L. experimental bounds on the BSM parameter $\amt$ ($\emt$) by IceCube~\cite{Aartsen:2017ibm, IceCube:2022ubv}. The significance of points P1$\ldots$P6 has been explained in the text.}
	\label{fig:2d}
\end{figure}
%=============================%

In this section, we explore the extent to which the LV vs. NSI discrimination is possible at the proposed 50 kt ICAL detector at INO. This detector would be sensitive to multi-GeV atmospheric $\nu_\mu$ and $\bar\nu_\mu$. It would detect muons in the energy range of 1 -- 25 GeV with the energy resolution of 10 -- 15\% and zenith angle resolution of $< 1^\circ$ for all but the almost-horizontal muons~\cite{Chatterjee:2014vta}. It would also measure the energy of hadron showers produced during atmospheric $\nu_\mu$ interaction events with a resolution of 35\%--70\%~\cite{Devi:2013wxa}. The magnetic field of 1.5 Tesla would give ICAL the unique capability of muon charge identification (CID), which in turn, would enable it to identify $\nu_\mu$ and $\bar\nu_\mu$ events separately with the CID efficiency of 98 -- 99\%, for muons beyond a few GeV to 50 GeV~\cite{Chatterjee:2014vta}. We simulate the events at the ICAL detector by using the NUANCE neutrino event generator~\cite{Casper:2002sd} with the atmospheric neutrino flux from Ref.~\cite{Honda:2015fha} to calculate the number of unoscillated events generated via the charged-current interactions. We incorporate the probabilities using the reweighting algorithm~\cite{Ghosh:2012px, Thakore:2013xqa}. We take into account the detector properties in the form of reconstruction efficiency, CID efficiency, and resolutions of muon energy, muon zenith angle and hadron energy, using the ICAL migration matrices~\cite{Chatterjee:2014vta, Devi:2013wxa}.

After an exposure of 1000 kt$\cdot$yr, about 8800 $\mu^-$ and 4000 $\mu^+$ reconstructed events are expected at ICAL. We use the reconstructed values of muon energy and zenith angle ($E_{\mu}^{\rm rec}$ and $\cos\theta_\mu^{\rm rec}$), along with hadron energy (${E'}_{\rm had}^{\rm rec}$), as observables~\cite{Devi:2014yaa}. The analysis is carried out using fine bins in $\cos\theta_\mu^{\rm rec}$ for core-passing  neutrinos~\cite{Upadhyay:2022jfd}. The data is simulated using nonzero $\amt^{\rm true}$, while the fit is attempted using nonzero $\emt^{\rm test}$. Using the frequentist approach, the median sensitivity of the detector to distinguish between LV and NSI scenarios, quantified by
\begin{align}
\Delta\chi^2 &= \chi^2\big(\amt^{\rm test} = 0,\,\emt^{\rm test}\big) - \chi^2\big(\amt^{\rm true},\,\,\emt^{\rm true} = 0\big) \;,
\label{Eq:4}
\end{align}
is calculated with the Poissonian $\chi^2$~\cite{Baker:1983tu, Blennow:2013oma} (see \ref{App:Stat} for more details). We restrict the values of $\amt$ and $\emt$ to be real. The five systematic uncertainties included are: 20\% error in flux normalization, 10\% error in cross section, 5\% energy dependent tilt error in flux, 5\% uncertainty on the zenith angle dependence of the flux, and 5\% overall systematics for both $\nu_\mu$ and $\bar\nu_\mu$ events~\cite{Kameda:2002fx, Gonzalez-Garcia:2004pka, Ghosh:2012px, Thakore:2013xqa, Devi:2014yaa}. The oscillation parameters values given earlier in this study are taken to be fixed, since by the time 1000 kt$\cdot$yr data is available, the oscillation parameters would have been measured quite precisely.
In Fig.~\ref{fig:chisq}, we show the best-fit values of $\emt^{\rm test}$ obtained for a range of $\amt^{\rm true}$ values, and the $\Delta\chi^2$ at which the hypothesis of NSI can be rejected. The figure clearly brings out the advantage that CID provides: it enables the discrimination at the level of $\Delta \chi^2 \approx 4$ for $|\amt| \ge 0.5 \times 10^{-23}$ GeV.

In Fig.~\ref{fig:2d}, we quantify the extent to which an NSI interpretation would mimic, or fail to mimic, an actual LV scenario. We scan over a range of $\emt^{\rm test}$ for each value of $\amt^{\rm true}$, and calculate the $\Delta\chi^2$ values. These values, in turn, indicate the confidence level (1 d.o.f.) at which the $\emt^{\rm test}$ would mimic a given value of $\amt^{\rm true}$. The results in Fig.~\ref{fig:2d}
can be interpreted as follows:
\begin{itemize}
	\item The points P1 and P2 lie on the line $\emt^{\rm test}$ = 0, i.e., the line which compares the SM against the LV scenarios. For these points, $\Delta\chi^2 = 3.84$. Thus, the points on this line to the left of P1 and the right of P2 indicate the values of $\amt^{\rm true}$ for which the SM would be disfavored to more than 95\% confidence level.
	\item The points P3 and P4 lie on the line $\amt^{\rm true}$ = 0, i.e., when there is no LV. For these points, $\Delta\chi^2 = 3.84$.	Thus, in the no-BSM scenario, the NSI parameter $\emt^{\rm test}$ would be resticted to the segment P3--P4 at 95\% confidence level.
	\item The green-dashed diagonal line approximately passes through the best-fit values of $\emt^{\rm test}$. This line corresponds to eq. (\ref{eq:degen}) with an average mass density of $\rho^{Earth}_{avg} \approx 5.5\, {\rm g/cm^3}$.
	\item The points P5 and P6 on the diagonal line represent the values of $\amt^{\rm true}$ beyond which any $\emt^{\rm test}$ would be ruled out to more than 95\% confidence level.
	\item The current experimental bounds on these BSM parameters from the IceCube experiment are $|\amt| \le 0.2 \times 10^{-23}$ GeV~\cite{Aartsen:2017ibm} and $-\,0.0123 \le \emt \le 0.0093$~\cite{IceCube:2022ubv} at 90\% C.L., considering these new physics scenarios one-at-a-time. The shaded regions depict these allowed ranges.
\end{itemize}
As can be inferred from the figure, 1000 kt$\cdot$yr of ICAL data can distinguish between LV and NSI scenarios with up to 95\% C.L. for $|\amt^{\rm true}| \le 0.5 \times 10^{-23}$ GeV. In the scenarios with $|\amt^{\rm true}| > 0.2\times 10^{-23}$ GeV, these data can distinguish between LV and NSI to more than 68\% confidence level. Higher exposure and the combination with data from other atmospheric neutrino experiments would enhance this sensitivity.

%=============================%
\section{Concluding remarks}
%=============================% 
Future high-precision neutrino experiments may reveal the presence of new physics scenarios beyond oscillations. However, it is not enough to identify the presence of BSM physics; it is essential to identify its source and nature. In this work, we have pointed out for the first time that a long-baseline neutrino oscillation experiment cannot distinguish between the two popular BSM scenarios of LV and NSI. We have argued that the key to resolving this degeneracy is the observation of neutrinos passing through the core of Earth, and hence atmospheric neutrino experiments are critical for this purpose. We demonstrate that the ICAL experiment, that has excellent energy and directional resolutions for muons and can identify the muon charge, may be able to distinguish between these two scenarios. Using the detailed and rigorous simulation tools developed by the INO collaboration, we further estimate the sensitivity of the proposed ICAL detector for discriminating between the LV and NSI scenarios with 1000 kt$\cdot$yr exposure.  

The analysis performed in this work can also be adapted for the currently running atmospheric neutrino experiments like Super-K, IceCube, DeepCore, ORCA, some of which also have sensitivity to electron events. The high-precision atmospheric neutrino data expected from the upcoming experiments like Hyper-K, DUNE and P-ONE will undoubtedly improve the prospects of the discrimination between LV and NSI scenarios.

%=============================%
\section*{Acknowledgements}
%=============================%
We thank the members of the INO-ICAL collaboration for their valuable comments and constructive inputs. We sincerely thank A. Raychaudhuri and S. Uma Sankar for their careful reading of the manuscript and for providing useful suggestions. We thank V. A. Kosteleck$\acute{\rm y}$, P. Denton, and S. Palomares-Ruiz for their valuable comments. We acknowledge the support from the Department of Atomic Energy (DAE), Govt. of India, under the Project Identification Numbers RTI4002 and RIO 4001. S.K.A. is supported by the Young  Scientist Project [INSA/SP/YSP/144/2017/1578] from the Indian National Science Academy (INSA). S.K.A. acknowledges the financial support from the Swarnajayanti Fellowship Research Grant (No. DST/SJF/PSA- 05/2019-20) provided by the Department of Science and Technology (DST), Govt. of India, and the Research Grant (File no. SB/SJF/2020-21/21) provided by the  Science and Engineering Research Board (SERB) under the Swarnajayanti Fellowship by the DST, Govt. of India. S.K.A would like to thank the United States-India Educational Foundation for providing the financial support through the Fulbright-Nehru Academic and Professional Excellence Fellowship (Award No. 2710/F-N APE/2021). The numerical simulations are performed using the high-performance computing facilities CCHPC-19 and Sim01 at the Tata Institute of Fundamental Research (TIFR), Mumbai, India.

\appendix

%=============================%
\section{Methodology of Statistical Analysis}
\label{App:Stat}
%=============================%
We calculate the median sensitivity of ICAL in terms of $\chi^2$, in a frequentist approach~\cite{Blennow:2013oma} for discriminating LV from NSI, assuming the Poissonian distribution~\cite{Baker:1983tu}. In our analysis, we define the $\chi^2$ for the reconstructed $\mu^-$ and $\mu^+$ events separately, by minimizing it over systematic uncertainties as follows: 
\begin{equation}
	\chi^2\left(\mu^\pm\right) = \mathop{\text{min}}_{\xi_l} \sum_{i=1}^{N_{{E'}_\text{had}^\text{rec}}} \sum_{j=1}^{N_{E_{\mu^\pm}^\text{rec}}} \sum_{k=1}^{N_{\cos\theta_{\mu^\pm}^\text{rec}}} 2\left[(N_{ijk}^\text{th} - N_{ijk}^\text{obs})\,+\,N_{ijk}^\text{obs} \ln\left(\frac{N_{ijk}^\text{obs} }{N_{ijk}^\text{th}}\right)\right] + \sum_{l = 1}^5 \xi_l^2
\label{App:Eq1}
\end{equation}
where 
\begin{equation}
	N_{ijk}^\text{th} = N_{ijk}^{0\; \text{th}}\left(1 + \sum_{l=1}^5 \pi^l_{ijk}\xi_l\right).
\label{App:Eq2}
\end{equation}
\noindent
Here, $N_{ijk}^\text{th}$ and $N_{ijk}^\text{obs}$ represent the number of expected and observed reconstructed $\mu^-$ and $\mu^+$ events for a given  $(E_\mu^\text{rec}, \cos\theta_\mu^\text{rec}, {E'}_\text{had}^\text{rec})$ bin, respectively. $N_{ijk}^{0\; \text{th}}$ corresponds to the theoretical prediction of reconstructed events. We adopt the pull method~\cite{GonzalezGarcia:2004wg,Huber:2002mx,Fogli:2002pt} to address the fluctuations in the theoretically predicted events due to the systematic uncertainties ($\pi^l_{ijk}$). The pull method allows us to parametrize the systematic and theoretical uncertainties in terms of a set of variables, the so-called pull-variables $\xi_l$. We use a linearized approximation while using the pull method to account the five systematic uncertainties, namely: the flux normalization error, uncertainties in cross sections, the energy-dependent tilt error in neutrino flux, uncertainties in the zenith angle dependence of the flux, and the error in overall systematics. The total $\chi^2$ is a sum of $\chi^2\left(\mu^-\right)$ and $\chi^2\left(\mu^+\right)$ as follows:
\begin{equation}
	\chi^2 = \chi^2\left(\mu^-\right) \, + \, \chi^2\left(\mu^+\right).
\label{App:Eq3}
\end{equation}

Note that the systematic uncertainties in neutrinos and antineutrinos are treated independently. Thus, there are overall 10 sources of systematic uncertainties that are taken care of in our analysis.

%=============================%
\bibliographystyle{elsarticle-num-names}

\begin{thebibliography}{124}
	\expandafter\ifx\csname natexlab\endcsname\relax\def\natexlab#1{#1}\fi
	\providecommand{\url}[1]{\texttt{#1}}
	\providecommand{\href}[2]{#2}
	\providecommand{\path}[1]{#1}
	\providecommand{\DOIprefix}{doi:}
	\providecommand{\ArXivprefix}{arXiv:}
	\providecommand{\URLprefix}{URL: }
	\providecommand{\Pubmedprefix}{pmid:}
	\providecommand{\doi}[1]{\href{http://dx.doi.org/#1}{\path{#1}}}
	\providecommand{\Pubmed}[1]{\href{pmid:#1}{\path{#1}}}
	\providecommand{\bibinfo}[2]{#2}
	\ifx\xfnm\relax \def\xfnm[#1]{\unskip,\space#1}\fi
	%Type = Article
	\bibitem[{Zyla et~al.(2020)}]{Zyla:2020zbs}
	\bibinfo{author}{P.~Zyla}, et~al. (\bibinfo{collaboration}{Particle Data
		Group}),
	\newblock \bibinfo{title}{{Review of Particle Physics}},
	\newblock \bibinfo{journal}{PTEP} \bibinfo{volume}{2020} (\bibinfo{year}{2020})
	\bibinfo{pages}{083C01}. \DOIprefix\doi{10.1093/ptep/ptaa104},
	\bibinfo{note}{and 2021 update}.
	%Type = Article
	\bibitem[{Fukuda et~al.(1999)}]{Super-Kamiokande:1998oic}
	\bibinfo{author}{Y.~Fukuda}, et~al.
	(\bibinfo{collaboration}{Super-Kamiokande}),
	\newblock \bibinfo{title}{{Constraints on neutrino oscillation parameters from
			the measurement of day night solar neutrino fluxes at Super-Kamiokande}},
	\newblock \bibinfo{journal}{Phys. Rev. Lett.} \bibinfo{volume}{82}
	(\bibinfo{year}{1999}) \bibinfo{pages}{1810--1814}.
	\DOIprefix\doi{10.1103/PhysRevLett.82.1810}.
	\href{http://arxiv.org/abs/hep-ex/9812009}{{\tt arXiv:hep-ex/9812009}}.
	%Type = Article
	\bibitem[{Fukuda et~al.(2001)}]{Super-Kamiokande:2001bfk}
	\bibinfo{author}{S.~Fukuda}, et~al.
	(\bibinfo{collaboration}{Super-Kamiokande}),
	\newblock \bibinfo{title}{{Constraints on neutrino oscillations using 1258 days
			of Super-Kamiokande solar neutrino data}},
	\newblock \bibinfo{journal}{Phys. Rev. Lett.} \bibinfo{volume}{86}
	(\bibinfo{year}{2001}) \bibinfo{pages}{5656--5660}.
	\DOIprefix\doi{10.1103/PhysRevLett.86.5656}.
	\href{http://arxiv.org/abs/hep-ex/0103033}{{\tt arXiv:hep-ex/0103033}}.
	%Type = Article
	\bibitem[{Ahmad et~al.(2001)}]{SNO:2001kpb}
	\bibinfo{author}{Q.~R. Ahmad}, et~al. (\bibinfo{collaboration}{SNO}),
	\newblock \bibinfo{title}{{Measurement of the rate of $\nu_e+d \to p+p+e^-$
			interactions produced by $^8$B solar neutrinos at the Sudbury Neutrino
			Observatory}},
	\newblock \bibinfo{journal}{Phys. Rev. Lett.} \bibinfo{volume}{87}
	(\bibinfo{year}{2001}) \bibinfo{pages}{071301}.
	\DOIprefix\doi{10.1103/PhysRevLett.87.071301}.
	\href{http://arxiv.org/abs/nucl-ex/0106015}{{\tt arXiv:nucl-ex/0106015}}.
	%Type = Article
	\bibitem[{Fukuda et~al.(2002)}]{Super-Kamiokande:2002ujc}
	\bibinfo{author}{S.~Fukuda}, et~al.
	(\bibinfo{collaboration}{Super-Kamiokande}),
	\newblock \bibinfo{title}{{Determination of solar neutrino oscillation
			parameters using 1496 days of Super-Kamiokande I data}},
	\newblock \bibinfo{journal}{Phys. Lett. B} \bibinfo{volume}{539}
	(\bibinfo{year}{2002}) \bibinfo{pages}{179--187}.
	\DOIprefix\doi{10.1016/S0370-2693(02)02090-7}.
	\href{http://arxiv.org/abs/hep-ex/0205075}{{\tt arXiv:hep-ex/0205075}}.
	%Type = Article
	\bibitem[{Hosaka et~al.(2006)}]{Super-Kamiokande:2005wtt}
	\bibinfo{author}{J.~Hosaka}, et~al.
	(\bibinfo{collaboration}{Super-Kamiokande}),
	\newblock \bibinfo{title}{{Solar neutrino measurements in super-Kamiokande-I}},
	\newblock \bibinfo{journal}{Phys. Rev. D} \bibinfo{volume}{73}
	(\bibinfo{year}{2006}) \bibinfo{pages}{112001}.
	\DOIprefix\doi{10.1103/PhysRevD.73.112001}.
	\href{http://arxiv.org/abs/hep-ex/0508053}{{\tt arXiv:hep-ex/0508053}}.
	%Type = Article
	\bibitem[{Cravens et~al.(2008)}]{Super-Kamiokande:2008ecj}
	\bibinfo{author}{J.~P. Cravens}, et~al.
	(\bibinfo{collaboration}{Super-Kamiokande}),
	\newblock \bibinfo{title}{{Solar neutrino measurements in
			Super-Kamiokande-II}},
	\newblock \bibinfo{journal}{Phys. Rev. D} \bibinfo{volume}{78}
	(\bibinfo{year}{2008}) \bibinfo{pages}{032002}.
	\DOIprefix\doi{10.1103/PhysRevD.78.032002}.
	\href{http://arxiv.org/abs/0803.4312}{{\tt arXiv:0803.4312}}.
	%Type = Article
	\bibitem[{Abe et~al.(2011)}]{Super-Kamiokande:2010tar}
	\bibinfo{author}{K.~Abe}, et~al. (\bibinfo{collaboration}{Super-Kamiokande}),
	\newblock \bibinfo{title}{{Solar neutrino results in Super-Kamiokande-III}},
	\newblock \bibinfo{journal}{Phys. Rev. D} \bibinfo{volume}{83}
	(\bibinfo{year}{2011}) \bibinfo{pages}{052010}.
	\DOIprefix\doi{10.1103/PhysRevD.83.052010}.
	\href{http://arxiv.org/abs/1010.0118}{{\tt arXiv:1010.0118}}.
	%Type = Article
	\bibitem[{Aharmim et~al.(2013)}]{SNO:2011hxd}
	\bibinfo{author}{B.~Aharmim}, et~al. (\bibinfo{collaboration}{SNO}),
	\newblock \bibinfo{title}{{Combined Analysis of all Three Phases of Solar
			Neutrino Data from the Sudbury Neutrino Observatory}},
	\newblock \bibinfo{journal}{Phys. Rev. C} \bibinfo{volume}{88}
	(\bibinfo{year}{2013}) \bibinfo{pages}{025501}.
	\DOIprefix\doi{10.1103/PhysRevC.88.025501}.
	\href{http://arxiv.org/abs/1109.0763}{{\tt arXiv:1109.0763}}.
	%Type = Article
	\bibitem[{Achar et~al.(1965)}]{Achar:1965ova}
	\bibinfo{author}{C.~V. Achar}, et~al.,
	\newblock \bibinfo{title}{{Detection of muons produced by cosmic ray neutrinos
			deep underground}},
	\newblock \bibinfo{journal}{Phys. Lett.} \bibinfo{volume}{18}
	(\bibinfo{year}{1965}) \bibinfo{pages}{196--199}.
	\DOIprefix\doi{10.1016/0031-9163(65)90712-2}.
	%Type = Article
	\bibitem[{Fukuda et~al.(1998)}]{Fukuda:1998mi}
	\bibinfo{author}{Y.~Fukuda}, et~al.
	(\bibinfo{collaboration}{Super-Kamiokande}),
	\newblock \bibinfo{title}{{Evidence for oscillation of atmospheric neutrinos}},
	\newblock \bibinfo{journal}{Phys. Rev. Lett.} \bibinfo{volume}{81}
	(\bibinfo{year}{1998}) \bibinfo{pages}{1562}.
	\DOIprefix\doi{10.1103/PhysRevLett.81.1562}.
	\href{http://arxiv.org/abs/hep-ex/9807003}{{\tt arXiv:hep-ex/9807003}}.
	%Type = Article
	\bibitem[{Ashie et~al.(2004)}]{Ashie2004}
	\bibinfo{author}{Y.~Ashie}, et~al. (\bibinfo{collaboration}{Super-Kamiokande}),
	\newblock \bibinfo{title}{{Evidence for an oscillatory signature in atmospheric
			neutrino oscillation}},
	\newblock \bibinfo{journal}{Phys. Rev. Lett.} \bibinfo{volume}{93}
	(\bibinfo{year}{2004}) \bibinfo{pages}{101801}.
	\DOIprefix\doi{10.1103/PhysRevLett.93.101801}.
	\href{http://arxiv.org/abs/hep-ex/0404034}{{\tt arXiv:hep-ex/0404034}}.
	%Type = Article
	\bibitem[{Aartsen et~al.(2015)}]{IceCube:2014flw}
	\bibinfo{author}{M.~G. Aartsen}, et~al. (\bibinfo{collaboration}{IceCube}),
	\newblock \bibinfo{title}{{Determining neutrino oscillation parameters from
			atmospheric muon neutrino disappearance with three years of IceCube DeepCore
			data}},
	\newblock \bibinfo{journal}{Phys. Rev. D} \bibinfo{volume}{91}
	(\bibinfo{year}{2015}) \bibinfo{pages}{072004}.
	\DOIprefix\doi{10.1103/PhysRevD.91.072004}.
	\href{http://arxiv.org/abs/1410.7227}{{\tt arXiv:1410.7227}}.
	%Type = Article
	\bibitem[{Abe et~al.(2018)}]{Super-Kamiokande:2017yvm}
	\bibinfo{author}{K.~Abe}, et~al. (\bibinfo{collaboration}{Super-Kamiokande}),
	\newblock \bibinfo{title}{{Atmospheric neutrino oscillation analysis with
			external constraints in Super-Kamiokande I-IV}},
	\newblock \bibinfo{journal}{Phys. Rev. D} \bibinfo{volume}{97}
	(\bibinfo{year}{2018}) \bibinfo{pages}{072001}.
	\DOIprefix\doi{10.1103/PhysRevD.97.072001}.
	\href{http://arxiv.org/abs/1710.09126}{{\tt arXiv:1710.09126}}.
	%Type = Article
	\bibitem[{Eguchi et~al.(2003)}]{Eguchi:2002dm}
	\bibinfo{author}{K.~Eguchi}, et~al. (\bibinfo{collaboration}{KamLAND}),
	\newblock \bibinfo{title}{{First results from KamLAND: Evidence for reactor
			anti-neutrino disappearance}},
	\newblock \bibinfo{journal}{Phys. Rev. Lett.} \bibinfo{volume}{90}
	(\bibinfo{year}{2003}) \bibinfo{pages}{021802}.
	\DOIprefix\doi{10.1103/PhysRevLett.90.021802}.
	\href{http://arxiv.org/abs/hep-ex/0212021}{{\tt arXiv:hep-ex/0212021}}.
	%Type = Article
	\bibitem[{Araki et~al.(2005)}]{Araki:2004mb}
	\bibinfo{author}{T.~Araki}, et~al. (\bibinfo{collaboration}{KamLAND
		Collaboration}),
	\newblock \bibinfo{title}{{Measurement of neutrino oscillation with KamLAND:
			Evidence of spectral distortion}},
	\newblock \bibinfo{journal}{Phys.Rev.Lett.} \bibinfo{volume}{94}
	(\bibinfo{year}{2005}) \bibinfo{pages}{081801}.
	\DOIprefix\doi{10.1103/PhysRevLett.94.081801}.
	\href{http://arxiv.org/abs/hep-ex/0406035}{{\tt arXiv:hep-ex/0406035}}.
	%Type = Article
	\bibitem[{An et~al.(2012)}]{An:2012eh}
	\bibinfo{author}{F.~P. An}, et~al. (\bibinfo{collaboration}{Daya Bay}),
	\newblock \bibinfo{title}{{Observation of electron-antineutrino disappearance
			at Daya Bay}},
	\newblock \bibinfo{journal}{Phys. Rev. Lett.} \bibinfo{volume}{108}
	(\bibinfo{year}{2012}) \bibinfo{pages}{171803}.
	\DOIprefix\doi{10.1103/PhysRevLett.108.171803}.
	\href{http://arxiv.org/abs/1203.1669}{{\tt arXiv:1203.1669}}.
	%Type = Article
	\bibitem[{Ahn et~al.(2012)}]{Ahn:2012nd}
	\bibinfo{author}{J.~K. Ahn}, et~al. (\bibinfo{collaboration}{RENO}),
	\newblock \bibinfo{title}{{Observation of Reactor Electron Antineutrino
			Disappearance in the RENO Experiment}},
	\newblock \bibinfo{journal}{Phys. Rev. Lett.} \bibinfo{volume}{108}
	(\bibinfo{year}{2012}) \bibinfo{pages}{191802}.
	\DOIprefix\doi{10.1103/PhysRevLett.108.191802}.
	\href{http://arxiv.org/abs/1204.0626}{{\tt arXiv:1204.0626}}.
	%Type = Article
	\bibitem[{Gando et~al.(2013)}]{KamLAND:2013rgu}
	\bibinfo{author}{A.~Gando}, et~al. (\bibinfo{collaboration}{KamLAND}),
	\newblock \bibinfo{title}{{Reactor On-Off Antineutrino Measurement with
			KamLAND}},
	\newblock \bibinfo{journal}{Phys. Rev. D} \bibinfo{volume}{88}
	(\bibinfo{year}{2013}) \bibinfo{pages}{033001}.
	\DOIprefix\doi{10.1103/PhysRevD.88.033001}.
	\href{http://arxiv.org/abs/1303.4667}{{\tt arXiv:1303.4667}}.
	%Type = Article
	\bibitem[{Bak et~al.(2018)}]{RENO:2018dro}
	\bibinfo{author}{G.~Bak}, et~al. (\bibinfo{collaboration}{RENO}),
	\newblock \bibinfo{title}{{Measurement of Reactor Antineutrino Oscillation
			Amplitude and Frequency at RENO}},
	\newblock \bibinfo{journal}{Phys. Rev. Lett.} \bibinfo{volume}{121}
	(\bibinfo{year}{2018}) \bibinfo{pages}{201801}.
	\DOIprefix\doi{10.1103/PhysRevLett.121.201801}.
	\href{http://arxiv.org/abs/1806.00248}{{\tt arXiv:1806.00248}}.
	%Type = Article
	\bibitem[{Adey et~al.(2018)}]{DayaBay:2018yms}
	\bibinfo{author}{D.~Adey}, et~al. (\bibinfo{collaboration}{Daya Bay}),
	\newblock \bibinfo{title}{{Measurement of the Electron Antineutrino Oscillation
			with 1958 Days of Operation at Daya Bay}},
	\newblock \bibinfo{journal}{Phys. Rev. Lett.} \bibinfo{volume}{121}
	(\bibinfo{year}{2018}) \bibinfo{pages}{241805}.
	\DOIprefix\doi{10.1103/PhysRevLett.121.241805}.
	\href{http://arxiv.org/abs/1809.02261}{{\tt arXiv:1809.02261}}.
	%Type = Article
	\bibitem[{de~Kerret et~al.(2020)}]{DoubleChooz:2019qbj}
	\bibinfo{author}{H.~de~Kerret}, et~al. (\bibinfo{collaboration}{Double Chooz}),
	\newblock \bibinfo{title}{{Double Chooz $\theta_{13}$ measurement via total
			neutron capture detection}},
	\newblock \bibinfo{journal}{Nature Phys.} \bibinfo{volume}{16}
	(\bibinfo{year}{2020}) \bibinfo{pages}{558--564}.
	\DOIprefix\doi{10.1038/s41567-020-0831-y}.
	\href{http://arxiv.org/abs/1901.09445}{{\tt arXiv:1901.09445}}.
	%Type = Article
	\bibitem[{Aliu et~al.(2005)}]{K2K:2004iot}
	\bibinfo{author}{E.~Aliu}, et~al. (\bibinfo{collaboration}{K2K}),
	\newblock \bibinfo{title}{{Evidence for muon neutrino oscillation in an
			accelerator-based experiment}},
	\newblock \bibinfo{journal}{Phys. Rev. Lett.} \bibinfo{volume}{94}
	(\bibinfo{year}{2005}) \bibinfo{pages}{081802}.
	\DOIprefix\doi{10.1103/PhysRevLett.94.081802}.
	\href{http://arxiv.org/abs/hep-ex/0411038}{{\tt arXiv:hep-ex/0411038}}.
	%Type = Article
	\bibitem[{Adamson et~al.(2008)}]{Adamson:2008zt}
	\bibinfo{author}{P.~Adamson}, et~al. (\bibinfo{collaboration}{MINOS}),
	\newblock \bibinfo{title}{{Measurement of Neutrino Oscillations with the MINOS
			Detectors in the NuMI Beam}},
	\newblock \bibinfo{journal}{Phys.Rev.Lett.} \bibinfo{volume}{101}
	(\bibinfo{year}{2008}) \bibinfo{pages}{131802}.
	\DOIprefix\doi{10.1103/PhysRevLett.101.131802}.
	\href{http://arxiv.org/abs/0806.2237}{{\tt arXiv:0806.2237}}.
	%Type = Article
	\bibitem[{Adamson et~al.(2013{\natexlab{a}})}]{MINOS:2013xrl}
	\bibinfo{author}{P.~Adamson}, et~al. (\bibinfo{collaboration}{MINOS}),
	\newblock \bibinfo{title}{{Electron neutrino and antineutrino appearance in the
			full MINOS data sample}},
	\newblock \bibinfo{journal}{Phys. Rev. Lett.} \bibinfo{volume}{110}
	(\bibinfo{year}{2013}{\natexlab{a}}) \bibinfo{pages}{171801}.
	\DOIprefix\doi{10.1103/PhysRevLett.110.171801}.
	\href{http://arxiv.org/abs/1301.4581}{{\tt arXiv:1301.4581}}.
	%Type = Article
	\bibitem[{Adamson et~al.(2013{\natexlab{b}})}]{MINOS:2013utc}
	\bibinfo{author}{P.~Adamson}, et~al. (\bibinfo{collaboration}{MINOS}),
	\newblock \bibinfo{title}{{Measurement of Neutrino and Antineutrino
			Oscillations Using Beam and Atmospheric Data in MINOS}},
	\newblock \bibinfo{journal}{Phys. Rev. Lett.} \bibinfo{volume}{110}
	(\bibinfo{year}{2013}{\natexlab{b}}) \bibinfo{pages}{251801}.
	\DOIprefix\doi{10.1103/PhysRevLett.110.251801}.
	\href{http://arxiv.org/abs/1304.6335}{{\tt arXiv:1304.6335}}.
	%Type = Article
	\bibitem[{Abe et~al.(2020)}]{T2K:2019bcf}
	\bibinfo{author}{K.~Abe}, et~al. (\bibinfo{collaboration}{T2K}),
	\newblock \bibinfo{title}{{Constraint on the matter\textendash{}antimatter
			symmetry-violating phase in neutrino oscillations}},
	\newblock \bibinfo{journal}{Nature} \bibinfo{volume}{580}
	(\bibinfo{year}{2020}) \bibinfo{pages}{339--344}.
	\DOIprefix\doi{10.1038/s41586-020-2177-0}.
	\href{http://arxiv.org/abs/1910.03887}{{\tt arXiv:1910.03887}},
	\bibinfo{note}{[Erratum: Nature 583, E16 (2020)]}.
	%Type = Article
	\bibitem[{Acero et~al.(2019)}]{NOvA:2019cyt}
	\bibinfo{author}{M.~A. Acero}, et~al. (\bibinfo{collaboration}{NOvA}),
	\newblock \bibinfo{title}{{First Measurement of Neutrino Oscillation Parameters
			using Neutrinos and Antineutrinos by NOvA}},
	\newblock \bibinfo{journal}{Phys. Rev. Lett.} \bibinfo{volume}{123}
	(\bibinfo{year}{2019}) \bibinfo{pages}{151803}.
	\DOIprefix\doi{10.1103/PhysRevLett.123.151803}.
	\href{http://arxiv.org/abs/1906.04907}{{\tt arXiv:1906.04907}}.
	%Type = Article
	\bibitem[{Arg\"uelles et~al.(2020)}]{Arguelles:2019xgp}
	\bibinfo{author}{C.~A. Arg\"uelles}, et~al.,
	\newblock \bibinfo{title}{{New opportunities at the next-generation neutrino
			experiments I: BSM neutrino physics and dark matter}},
	\newblock \bibinfo{journal}{Rept. Prog. Phys.} \bibinfo{volume}{83}
	(\bibinfo{year}{2020}) \bibinfo{pages}{124201}.
	\DOIprefix\doi{10.1088/1361-6633/ab9d12}.
	\href{http://arxiv.org/abs/1907.08311}{{\tt arXiv:1907.08311}}.
	%Type = Article
	\bibitem[{Colladay and Kostelecky(1997)}]{Colladay:1996iz}
	\bibinfo{author}{D.~Colladay}, \bibinfo{author}{V.~A. Kostelecky},
	\newblock \bibinfo{title}{{CPT violation and the standard model}},
	\newblock \bibinfo{journal}{Phys. Rev. D} \bibinfo{volume}{55}
	(\bibinfo{year}{1997}) \bibinfo{pages}{6760--6774}.
	\DOIprefix\doi{10.1103/PhysRevD.55.6760}.
	\href{http://arxiv.org/abs/hep-ph/9703464}{{\tt arXiv:hep-ph/9703464}}.
	%Type = Article
	\bibitem[{Auerbach et~al.(2005)}]{Auerbach:2005tq}
	\bibinfo{author}{L.~B. Auerbach}, et~al. (\bibinfo{collaboration}{LSND}),
	\newblock \bibinfo{title}{{Tests of Lorentz violation in anti-nu(mu)
			---\ensuremath{>} anti-nu(e) oscillations}},
	\newblock \bibinfo{journal}{Phys. Rev. D} \bibinfo{volume}{72}
	(\bibinfo{year}{2005}) \bibinfo{pages}{076004}.
	\DOIprefix\doi{10.1103/PhysRevD.72.076004}.
	\href{http://arxiv.org/abs/hep-ex/0506067}{{\tt arXiv:hep-ex/0506067}}.
	%Type = Article
	\bibitem[{Adamson et~al.(2008)}]{Adamson:2008aa}
	\bibinfo{author}{P.~Adamson}, et~al. (\bibinfo{collaboration}{MINOS}),
	\newblock \bibinfo{title}{{Testing Lorentz Invariance and CPT Conservation with
			NuMI Neutrinos in the MINOS Near Detector}},
	\newblock \bibinfo{journal}{Phys. Rev. Lett.} \bibinfo{volume}{101}
	(\bibinfo{year}{2008}) \bibinfo{pages}{151601}.
	\DOIprefix\doi{10.1103/PhysRevLett.101.151601}.
	\href{http://arxiv.org/abs/0806.4945}{{\tt arXiv:0806.4945}}.
	%Type = Article
	\bibitem[{Adamson et~al.(2010)}]{Adamson:2010rn}
	\bibinfo{author}{P.~Adamson}, et~al. (\bibinfo{collaboration}{MINOS}),
	\newblock \bibinfo{title}{{A Search for Lorentz Invariance and CPT Violation
			with the MINOS Far Detector}},
	\newblock \bibinfo{journal}{Phys. Rev. Lett.} \bibinfo{volume}{105}
	(\bibinfo{year}{2010}) \bibinfo{pages}{151601}.
	\DOIprefix\doi{10.1103/PhysRevLett.105.151601}.
	\href{http://arxiv.org/abs/1007.2791}{{\tt arXiv:1007.2791}}.
	%Type = Article
	\bibitem[{Abbasi et~al.(2010)}]{Abbasi:2010kx}
	\bibinfo{author}{R.~Abbasi}, et~al. (\bibinfo{collaboration}{IceCube}),
	\newblock \bibinfo{title}{{Search for a Lorentz-violating sidereal signal with
			atmospheric neutrinos in IceCube}},
	\newblock \bibinfo{journal}{Phys. Rev. D} \bibinfo{volume}{82}
	(\bibinfo{year}{2010}) \bibinfo{pages}{112003}.
	\DOIprefix\doi{10.1103/PhysRevD.82.112003}.
	\href{http://arxiv.org/abs/1010.4096}{{\tt arXiv:1010.4096}}.
	%Type = Article
	\bibitem[{Aguilar-Arevalo et~al.(2013)}]{AguilarArevalo:2011yi}
	\bibinfo{author}{A.~A. Aguilar-Arevalo}, et~al.
	(\bibinfo{collaboration}{MiniBooNE}),
	\newblock \bibinfo{title}{{Test of Lorentz and CPT violation with Short
			Baseline Neutrino Oscillation Excesses}},
	\newblock \bibinfo{journal}{Phys. Lett. B} \bibinfo{volume}{718}
	(\bibinfo{year}{2013}) \bibinfo{pages}{1303--1308}.
	\DOIprefix\doi{10.1016/j.physletb.2012.12.020}.
	\href{http://arxiv.org/abs/1109.3480}{{\tt arXiv:1109.3480}}.
	%Type = Article
	\bibitem[{Kosteleck\'y and Russell(2011)}]{Kostelecky:2008ts}
	\bibinfo{author}{V.~A. Kosteleck\'y}, \bibinfo{author}{N.~Russell},
	\newblock \bibinfo{title}{Data tables for lorentz and $cpt$ violation},
	\newblock \bibinfo{journal}{Rev. Mod. Phys.} \bibinfo{volume}{83}
	(\bibinfo{year}{2011}) \bibinfo{pages}{11--31}. \URLprefix
	\url{https://link.aps.org/doi/10.1103/RevModPhys.83.11}.
	\DOIprefix\doi{10.1103/RevModPhys.83.11}.
	%Type = Article
	\bibitem[{Adamson et~al.(2012)}]{Adamson:2012hp}
	\bibinfo{author}{P.~Adamson}, et~al. (\bibinfo{collaboration}{MINOS}),
	\newblock \bibinfo{title}{{Search for Lorentz invariance and CPT violation with
			muon antineutrinos in the MINOS Near Detector}},
	\newblock \bibinfo{journal}{Phys. Rev. D} \bibinfo{volume}{85}
	(\bibinfo{year}{2012}) \bibinfo{pages}{031101}.
	\DOIprefix\doi{10.1103/PhysRevD.85.031101}.
	\href{http://arxiv.org/abs/1201.2631}{{\tt arXiv:1201.2631}}.
	%Type = Article
	\bibitem[{Abe et~al.(2012)}]{Abe:2012gw}
	\bibinfo{author}{Y.~Abe}, et~al. (\bibinfo{collaboration}{Double Chooz}),
	\newblock \bibinfo{title}{{First Test of Lorentz Violation with a Reactor-based
			Antineutrino Experiment}},
	\newblock \bibinfo{journal}{Phys. Rev. D} \bibinfo{volume}{86}
	(\bibinfo{year}{2012}) \bibinfo{pages}{112009}.
	\DOIprefix\doi{10.1103/PhysRevD.86.112009}.
	\href{http://arxiv.org/abs/1209.5810}{{\tt arXiv:1209.5810}}.
	%Type = Article
	\bibitem[{Rebel and Mufson(2013)}]{Rebel:2013vc}
	\bibinfo{author}{B.~Rebel}, \bibinfo{author}{S.~Mufson},
	\newblock \bibinfo{title}{{The Search for Neutrino-Antineutrino Mixing
			Resulting from Lorentz Invariance Violation using neutrino interactions in
			MINOS}},
	\newblock \bibinfo{journal}{Astropart. Phys.} \bibinfo{volume}{48}
	(\bibinfo{year}{2013}) \bibinfo{pages}{78--81}.
	\DOIprefix\doi{10.1016/j.astropartphys.2013.07.006}.
	\href{http://arxiv.org/abs/1301.4684}{{\tt arXiv:1301.4684}}.
	%Type = Article
	\bibitem[{D\'\i{}az et~al.(2013)D\'\i{}az, Katori, Spitz, and
		Conrad}]{Diaz:2013iba}
	\bibinfo{author}{J.~S. D\'\i{}az}, \bibinfo{author}{T.~Katori},
	\bibinfo{author}{J.~Spitz}, \bibinfo{author}{J.~M. Conrad},
	\newblock \bibinfo{title}{{Search for neutrino-antineutrino oscillations with a
			reactor experiment}},
	\newblock \bibinfo{journal}{Phys. Lett. B} \bibinfo{volume}{727}
	(\bibinfo{year}{2013}) \bibinfo{pages}{412--416}.
	\DOIprefix\doi{10.1016/j.physletb.2013.10.058}.
	\href{http://arxiv.org/abs/1307.5789}{{\tt arXiv:1307.5789}}.
	%Type = Article
	\bibitem[{Diaz and Schwetz(2016)}]{Diaz:2016fqd}
	\bibinfo{author}{J.~S. Diaz}, \bibinfo{author}{T.~Schwetz},
	\newblock \bibinfo{title}{{Limits on CPT violation from solar neutrinos}},
	\newblock \bibinfo{journal}{Phys. Rev. D} \bibinfo{volume}{93}
	(\bibinfo{year}{2016}) \bibinfo{pages}{093004}.
	\DOIprefix\doi{10.1103/PhysRevD.93.093004}.
	\href{http://arxiv.org/abs/1603.04468}{{\tt arXiv:1603.04468}}.
	%Type = Article
	\bibitem[{Abe et~al.(2015)}]{Abe:2014wla}
	\bibinfo{author}{K.~Abe}, et~al. (\bibinfo{collaboration}{Super-Kamiokande}),
	\newblock \bibinfo{title}{{Test of Lorentz invariance with atmospheric
			neutrinos}},
	\newblock \bibinfo{journal}{Phys. Rev. D} \bibinfo{volume}{91}
	(\bibinfo{year}{2015}) \bibinfo{pages}{052003}.
	\DOIprefix\doi{10.1103/PhysRevD.91.052003}.
	\href{http://arxiv.org/abs/1410.4267}{{\tt arXiv:1410.4267}}.
	%Type = Article
	\bibitem[{Abe et~al.(2017)}]{Abe:2017eot}
	\bibinfo{author}{K.~Abe}, et~al. (\bibinfo{collaboration}{T2K}),
	\newblock \bibinfo{title}{{Search for Lorentz and CPT violation using sidereal
			time dependence of neutrino flavor transitions over a short baseline}},
	\newblock \bibinfo{journal}{Phys. Rev. D} \bibinfo{volume}{95}
	(\bibinfo{year}{2017}) \bibinfo{pages}{111101}.
	\DOIprefix\doi{10.1103/PhysRevD.95.111101}.
	\href{http://arxiv.org/abs/1703.01361}{{\tt arXiv:1703.01361}}.
	%Type = Article
	\bibitem[{Adey et~al.(2018)}]{Adey:2018qsd}
	\bibinfo{author}{D.~Adey}, et~al. (\bibinfo{collaboration}{Daya Bay}),
	\newblock \bibinfo{title}{{Search for a time-varying electron antineutrino
			signal at Daya Bay}},
	\newblock \bibinfo{journal}{Phys. Rev. D} \bibinfo{volume}{98}
	(\bibinfo{year}{2018}) \bibinfo{pages}{092013}.
	\DOIprefix\doi{10.1103/PhysRevD.98.092013}.
	\href{http://arxiv.org/abs/1809.04660}{{\tt arXiv:1809.04660}}.
	%Type = Article
	\bibitem[{Aartsen et~al.(2018)}]{Aartsen:2017ibm}
	\bibinfo{author}{M.~G. Aartsen}, et~al. (\bibinfo{collaboration}{IceCube}),
	\newblock \bibinfo{title}{{Neutrino Interferometry for High-Precision Tests of
			Lorentz Symmetry with IceCube}},
	\newblock \bibinfo{journal}{Nature Phys.} \bibinfo{volume}{14}
	(\bibinfo{year}{2018}) \bibinfo{pages}{961--966}.
	\DOIprefix\doi{10.1038/s41567-018-0172-2}.
	\href{http://arxiv.org/abs/1709.03434}{{\tt arXiv:1709.03434}}.
	%Type = Article
	\bibitem[{Sahoo et~al.(2022)Sahoo, Kumar, and Agarwalla}]{Sahoo:2021dit}
	\bibinfo{author}{S.~Sahoo}, \bibinfo{author}{A.~Kumar}, \bibinfo{author}{S.~K.
		Agarwalla},
	\newblock \bibinfo{title}{{Probing Lorentz Invariance Violation with
			atmospheric neutrinos at INO-ICAL}},
	\newblock \bibinfo{journal}{JHEP} \bibinfo{volume}{03} (\bibinfo{year}{2022})
	\bibinfo{pages}{050}. \DOIprefix\doi{10.1007/JHEP03(2022)050}.
	\href{http://arxiv.org/abs/2110.13207}{{\tt arXiv:2110.13207}}.
	%Type = Article
	\bibitem[{Wolfenstein(1978)}]{Wolfenstein:1977nu}
	\bibinfo{author}{L.~Wolfenstein},
	\newblock \bibinfo{title}{Neutrino oscillations in matter},
	\newblock \bibinfo{journal}{Phys. Rev. D} \bibinfo{volume}{17}
	(\bibinfo{year}{1978}) \bibinfo{pages}{2369--2374}. \URLprefix
	\url{https://link.aps.org/doi/10.1103/PhysRevD.17.2369}.
	\DOIprefix\doi{10.1103/PhysRevD.17.2369}.
	%Type = Article
	\bibitem[{Ohlsson(2013)}]{Ohlsson:2012kf}
	\bibinfo{author}{T.~Ohlsson},
	\newblock \bibinfo{title}{{Status of non-standard neutrino interactions}},
	\newblock \bibinfo{journal}{Rept. Prog. Phys.} \bibinfo{volume}{76}
	(\bibinfo{year}{2013}) \bibinfo{pages}{044201}.
	\DOIprefix\doi{10.1088/0034-4885/76/4/044201}.
	\href{http://arxiv.org/abs/1209.2710}{{\tt arXiv:1209.2710}}.
	%Type = Article
	\bibitem[{Fornengo et~al.(2002)Fornengo, Maltoni, Tomas, and
		Valle}]{Fornengo:2001pm}
	\bibinfo{author}{N.~Fornengo}, \bibinfo{author}{M.~Maltoni},
	\bibinfo{author}{R.~Tomas}, \bibinfo{author}{J.~W.~F. Valle},
	\newblock \bibinfo{title}{{Probing neutrino nonstandard interactions with
			atmospheric neutrino data}},
	\newblock \bibinfo{journal}{Phys. Rev. D} \bibinfo{volume}{65}
	(\bibinfo{year}{2002}) \bibinfo{pages}{013010}.
	\DOIprefix\doi{10.1103/PhysRevD.65.013010}.
	\href{http://arxiv.org/abs/hep-ph/0108043}{{\tt arXiv:hep-ph/0108043}}.
	%Type = Article
	\bibitem[{Kopp et~al.(2010)Kopp, Machado, and Parke}]{Kopp:2010qt}
	\bibinfo{author}{J.~Kopp}, \bibinfo{author}{P.~A.~N. Machado},
	\bibinfo{author}{S.~J. Parke},
	\newblock \bibinfo{title}{{Interpretation of MINOS Data in Terms of
			Non-Standard Neutrino Interactions}},
	\newblock \bibinfo{journal}{Phys. Rev. D} \bibinfo{volume}{82}
	(\bibinfo{year}{2010}) \bibinfo{pages}{113002}.
	\DOIprefix\doi{10.1103/PhysRevD.82.113002}.
	\href{http://arxiv.org/abs/1009.0014}{{\tt arXiv:1009.0014}}.
	%Type = Article
	\bibitem[{Mitsuka et~al.(2011)}]{Super-Kamiokande:2011dam}
	\bibinfo{author}{G.~Mitsuka}, et~al.
	(\bibinfo{collaboration}{Super-Kamiokande}),
	\newblock \bibinfo{title}{{Study of Non-Standard Neutrino Interactions with
			Atmospheric Neutrino Data in Super-Kamiokande I and II}},
	\newblock \bibinfo{journal}{Phys. Rev. D} \bibinfo{volume}{84}
	(\bibinfo{year}{2011}) \bibinfo{pages}{113008}.
	\DOIprefix\doi{10.1103/PhysRevD.84.113008}.
	\href{http://arxiv.org/abs/1109.1889}{{\tt arXiv:1109.1889}}.
	%Type = Article
	\bibitem[{Choubey et~al.(2015)Choubey, Ghosh, Ohlsson, and
		Tiwari}]{Choubey:2015xha}
	\bibinfo{author}{S.~Choubey}, \bibinfo{author}{A.~Ghosh},
	\bibinfo{author}{T.~Ohlsson}, \bibinfo{author}{D.~Tiwari},
	\newblock \bibinfo{title}{{Neutrino Physics with Non-Standard Interactions at
			INO}},
	\newblock \bibinfo{journal}{JHEP} \bibinfo{volume}{12} (\bibinfo{year}{2015})
	\bibinfo{pages}{126}. \DOIprefix\doi{10.1007/JHEP12(2015)126}.
	\href{http://arxiv.org/abs/1507.02211}{{\tt arXiv:1507.02211}}.
	%Type = Article
	\bibitem[{Salvado et~al.(2017)Salvado, Mena, Palomares-Ruiz, and
		Rius}]{Salvado:2016uqu}
	\bibinfo{author}{J.~Salvado}, \bibinfo{author}{O.~Mena},
	\bibinfo{author}{S.~Palomares-Ruiz}, \bibinfo{author}{N.~Rius},
	\newblock \bibinfo{title}{{Non-standard interactions with high-energy
			atmospheric neutrinos at IceCube}},
	\newblock \bibinfo{journal}{JHEP} \bibinfo{volume}{01} (\bibinfo{year}{2017})
	\bibinfo{pages}{141}. \DOIprefix\doi{10.1007/JHEP01(2017)141}.
	\href{http://arxiv.org/abs/1609.03450}{{\tt arXiv:1609.03450}}.
	%Type = Article
	\bibitem[{Aartsen et~al.(2018)}]{IceCube:2017zcu}
	\bibinfo{author}{M.~G. Aartsen}, et~al. (\bibinfo{collaboration}{IceCube}),
	\newblock \bibinfo{title}{{Search for Nonstandard Neutrino Interactions with
			IceCube DeepCore}},
	\newblock \bibinfo{journal}{Phys. Rev. D} \bibinfo{volume}{97}
	(\bibinfo{year}{2018}) \bibinfo{pages}{072009}.
	\DOIprefix\doi{10.1103/PhysRevD.97.072009}.
	\href{http://arxiv.org/abs/1709.07079}{{\tt arXiv:1709.07079}}.
	%Type = Article
	\bibitem[{Farzan and Tortola(2018)}]{Farzan:2017xzy}
	\bibinfo{author}{Y.~Farzan}, \bibinfo{author}{M.~Tortola},
	\newblock \bibinfo{title}{{Neutrino oscillations and Non-Standard
			Interactions}},
	\newblock \bibinfo{journal}{Front.in Phys.} \bibinfo{volume}{6}
	(\bibinfo{year}{2018}) \bibinfo{pages}{10}.
	\DOIprefix\doi{10.3389/fphy.2018.00010}.
	\href{http://arxiv.org/abs/1710.09360}{{\tt arXiv:1710.09360}}.
	%Type = Article
	\bibitem[{Khatun et~al.(2020)Khatun, Chatterjee, Thakore, and
		Agarwalla}]{Khatun:2019tad}
	\bibinfo{author}{A.~Khatun}, \bibinfo{author}{S.~S. Chatterjee},
	\bibinfo{author}{T.~Thakore}, \bibinfo{author}{S.~K. Agarwalla},
	\newblock \bibinfo{title}{{Enhancing sensitivity to non-standard neutrino
			interactions at INO combining muon and hadron information}},
	\newblock \bibinfo{journal}{Eur. Phys. J. C} \bibinfo{volume}{80}
	(\bibinfo{year}{2020}) \bibinfo{pages}{533}.
	\DOIprefix\doi{10.1140/epjc/s10052-020-8097-1}.
	\href{http://arxiv.org/abs/1907.02027}{{\tt arXiv:1907.02027}}.
	%Type = Article
	\bibitem[{Kumar et~al.(2021)Kumar, Khatun, Agarwalla, and
		Dighe}]{Kumar:2021lrn}
	\bibinfo{author}{A.~Kumar}, \bibinfo{author}{A.~Khatun}, \bibinfo{author}{S.~K.
		Agarwalla}, \bibinfo{author}{A.~Dighe},
	\newblock \bibinfo{title}{{A New Approach to Probe Non-Standard Interactions in
			Atmospheric Neutrino Experiments}},
	\newblock \bibinfo{journal}{JHEP} \bibinfo{volume}{04} (\bibinfo{year}{2021})
	\bibinfo{pages}{159}. \DOIprefix\doi{10.1007/JHEP04(2021)159}.
	\href{http://arxiv.org/abs/2101.02607}{{\tt arXiv:2101.02607}}.
	%Type = Inproceedings
	\bibitem[{Hern\'andez~Rey et~al.(2021)Hern\'andez~Rey, Khan~Chowdhury, Manczak,
		Navas, and Zornoza}]{HernandezRey:2021qac}
	\bibinfo{author}{J.~J. Hern\'andez~Rey}, \bibinfo{author}{N.~R.
		Khan~Chowdhury}, \bibinfo{author}{J.~Manczak}, \bibinfo{author}{S.~Navas},
	\bibinfo{author}{J.~D. Zornoza} (\bibinfo{collaboration}{ANTARES, KM3NeT}),
	\newblock \bibinfo{title}{{Search for neutrino non-standard interactions with
			ANTARES and KM3NeT-ORCA}},
	\newblock in: \bibinfo{booktitle}{{9th Very Large Volume Neutrino Telescopes
			Workshop 2021}}, \bibinfo{year}{2021}.
	\href{http://arxiv.org/abs/2107.14296}{{\tt arXiv:2107.14296}}.
	%Type = Article
	\bibitem[{Abbasi et~al.(2021)}]{IceCube:2021euf}
	\bibinfo{author}{R.~Abbasi}, et~al. (\bibinfo{collaboration}{(IceCube
		Collaboration)*, IceCube}),
	\newblock \bibinfo{title}{{All-flavor constraints on nonstandard neutrino
			interactions and generalized matter potential with three years of IceCube
			DeepCore data}},
	\newblock \bibinfo{journal}{Phys. Rev. D} \bibinfo{volume}{104}
	(\bibinfo{year}{2021}) \bibinfo{pages}{072006}.
	\DOIprefix\doi{10.1103/PhysRevD.104.072006}.
	\href{http://arxiv.org/abs/2106.07755}{{\tt arXiv:2106.07755}}.
	%Type = Article
	\bibitem[{Albert et~al.(2021)}]{ANTARES:2021crm}
	\bibinfo{author}{A.~Albert}, et~al. (\bibinfo{collaboration}{ANTARES}),
	\newblock \bibinfo{title}{{Search for non-standard neutrino interactions with
			10 years of ANTARES data}}  (\bibinfo{year}{2021}).
	\href{http://arxiv.org/abs/2112.14517}{{\tt arXiv:2112.14517}}.
	%Type = Article
	\bibitem[{Denton and Pestes(2021)}]{Denton:2021rgt}
	\bibinfo{author}{P.~B. Denton}, \bibinfo{author}{R.~Pestes},
	\newblock \bibinfo{title}{{Neutrino oscillations through the
			Earth\textquoteright{}s core}},
	\newblock \bibinfo{journal}{Phys. Rev. D} \bibinfo{volume}{104}
	(\bibinfo{year}{2021}) \bibinfo{pages}{113007}.
	\DOIprefix\doi{10.1103/PhysRevD.104.113007}.
	\href{http://arxiv.org/abs/2110.01148}{{\tt arXiv:2110.01148}}.
	%Type = Article
	\bibitem[{Ahmed et~al.(2017)}]{ICAL:2015stm}
	\bibinfo{author}{S.~Ahmed}, et~al. (\bibinfo{collaboration}{ICAL}),
	\newblock \bibinfo{title}{{Physics Potential of the ICAL detector at the
			India-based Neutrino Observatory (INO)}},
	\newblock \bibinfo{journal}{Pramana} \bibinfo{volume}{88}
	(\bibinfo{year}{2017}) \bibinfo{pages}{79}.
	\DOIprefix\doi{10.1007/s12043-017-1373-4}.
	\href{http://arxiv.org/abs/1505.07380}{{\tt arXiv:1505.07380}}.
	%Type = Book
	\bibitem[{Polyakov(1987)}]{Polyakov:1987ez}
	\bibinfo{author}{A.~M. Polyakov}, \bibinfo{title}{{Gauge Fields and Strings}},
	volume~\bibinfo{volume}{3} of \textit{\bibinfo{series}{{Contemporary concepts
				in physics}}}, \bibinfo{publisher}{{Harwood Academic Publishers}},
	\bibinfo{year}{1987}.
	%Type = Article
	\bibitem[{Kostelecky and Samuel(1989{\natexlab{a}})}]{Kostelecky:1988zi}
	\bibinfo{author}{V.~A. Kostelecky}, \bibinfo{author}{S.~Samuel},
	\newblock \bibinfo{title}{{Spontaneous Breaking of Lorentz Symmetry in String
			Theory}},
	\newblock \bibinfo{journal}{Phys. Rev. D} \bibinfo{volume}{39}
	(\bibinfo{year}{1989}{\natexlab{a}}) \bibinfo{pages}{683}.
	\DOIprefix\doi{10.1103/PhysRevD.39.683}.
	%Type = Article
	\bibitem[{Kostelecky and Samuel(1989{\natexlab{b}})}]{Kostelecky:1989jp}
	\bibinfo{author}{V.~A. Kostelecky}, \bibinfo{author}{S.~Samuel},
	\newblock \bibinfo{title}{{Phenomenological Gravitational Constraints on
			Strings and Higher Dimensional Theories}},
	\newblock \bibinfo{journal}{Phys. Rev. Lett.} \bibinfo{volume}{63}
	(\bibinfo{year}{1989}{\natexlab{b}}) \bibinfo{pages}{224}.
	\DOIprefix\doi{10.1103/PhysRevLett.63.224}.
	%Type = Article
	\bibitem[{Kosteleck\'y and Samuel(1991)}]{Kostelecky:1990pe}
	\bibinfo{author}{V.~A. Kosteleck\'y}, \bibinfo{author}{S.~Samuel},
	\newblock \bibinfo{title}{Photon and graviton masses in string theories},
	\newblock \bibinfo{journal}{Phys. Rev. Lett.} \bibinfo{volume}{66}
	(\bibinfo{year}{1991}) \bibinfo{pages}{1811--1814}. \URLprefix
	\url{https://link.aps.org/doi/10.1103/PhysRevLett.66.1811}.
	\DOIprefix\doi{10.1103/PhysRevLett.66.1811}.
	%Type = Article
	\bibitem[{Kostelecky and Potting(1991)}]{Kostelecky:1991ak}
	\bibinfo{author}{V.~A. Kostelecky}, \bibinfo{author}{R.~Potting},
	\newblock \bibinfo{title}{{CPT and strings}},
	\newblock \bibinfo{journal}{Nucl. Phys. B} \bibinfo{volume}{359}
	(\bibinfo{year}{1991}) \bibinfo{pages}{545--570}.
	\DOIprefix\doi{10.1016/0550-3213(91)90071-5}.
	%Type = Article
	\bibitem[{Kostelecky and Potting(1996)}]{Kostelecky:1995qk}
	\bibinfo{author}{V.~A. Kostelecky}, \bibinfo{author}{R.~Potting},
	\newblock \bibinfo{title}{{Expectation values, Lorentz invariance, and CPT in
			the open bosonic string}},
	\newblock \bibinfo{journal}{Phys. Lett. B} \bibinfo{volume}{381}
	(\bibinfo{year}{1996}) \bibinfo{pages}{89--96}.
	\DOIprefix\doi{10.1016/0370-2693(96)00589-8}.
	\href{http://arxiv.org/abs/hep-th/9605088}{{\tt arXiv:hep-th/9605088}}.
	%Type = Article
	\bibitem[{Kostelecky et~al.(2000)Kostelecky, Perry, and
		Potting}]{Kostelecky:1999mu}
	\bibinfo{author}{V.~A. Kostelecky}, \bibinfo{author}{M.~Perry},
	\bibinfo{author}{R.~Potting},
	\newblock \bibinfo{title}{{Off-shell structure of the string sigma model}},
	\newblock \bibinfo{journal}{Phys. Rev. Lett.} \bibinfo{volume}{84}
	(\bibinfo{year}{2000}) \bibinfo{pages}{4541--4544}.
	\DOIprefix\doi{10.1103/PhysRevLett.84.4541}.
	\href{http://arxiv.org/abs/hep-th/9912243}{{\tt arXiv:hep-th/9912243}}.
	%Type = Article
	\bibitem[{Kostelecky and Potting(2001)}]{Kostelecky:2000hz}
	\bibinfo{author}{V.~A. Kostelecky}, \bibinfo{author}{R.~Potting},
	\newblock \bibinfo{title}{{Analytical construction of a nonperturbative vacuum
			for the open bosonic string}},
	\newblock \bibinfo{journal}{Phys. Rev. D} \bibinfo{volume}{63}
	(\bibinfo{year}{2001}) \bibinfo{pages}{046007}.
	\DOIprefix\doi{10.1103/PhysRevD.63.046007}.
	\href{http://arxiv.org/abs/hep-th/0008252}{{\tt arXiv:hep-th/0008252}}.
	%Type = Article
	\bibitem[{Gambini and Pullin(1999)}]{Gambini:1998it}
	\bibinfo{author}{R.~Gambini}, \bibinfo{author}{J.~Pullin},
	\newblock \bibinfo{title}{{Nonstandard optics from quantum space-time}},
	\newblock \bibinfo{journal}{Phys. Rev. D} \bibinfo{volume}{59}
	(\bibinfo{year}{1999}) \bibinfo{pages}{124021}.
	\DOIprefix\doi{10.1103/PhysRevD.59.124021}.
	\href{http://arxiv.org/abs/gr-qc/9809038}{{\tt arXiv:gr-qc/9809038}}.
	%Type = Article
	\bibitem[{Alfaro et~al.(2002)Alfaro, Morales-Tecotl, and
		Urrutia}]{Alfaro:2002xz}
	\bibinfo{author}{J.~Alfaro}, \bibinfo{author}{H.~A. Morales-Tecotl},
	\bibinfo{author}{L.~F. Urrutia},
	\newblock \bibinfo{title}{{Quantum gravity and spin 1/2 particles effective
			dynamics}},
	\newblock \bibinfo{journal}{Phys. Rev. D} \bibinfo{volume}{66}
	(\bibinfo{year}{2002}) \bibinfo{pages}{124006}.
	\DOIprefix\doi{10.1103/PhysRevD.66.124006}.
	\href{http://arxiv.org/abs/hep-th/0208192}{{\tt arXiv:hep-th/0208192}}.
	%Type = Article
	\bibitem[{Sudarsky et~al.(2002)Sudarsky, Urrutia, and
		Vucetich}]{Sudarsky:2002ue}
	\bibinfo{author}{D.~Sudarsky}, \bibinfo{author}{L.~Urrutia},
	\bibinfo{author}{H.~Vucetich},
	\newblock \bibinfo{title}{{New observational bounds to quantum gravity
			signals}},
	\newblock \bibinfo{journal}{Phys. Rev. Lett.} \bibinfo{volume}{89}
	(\bibinfo{year}{2002}) \bibinfo{pages}{231301}.
	\DOIprefix\doi{10.1103/PhysRevLett.89.231301}.
	\href{http://arxiv.org/abs/gr-qc/0204027}{{\tt arXiv:gr-qc/0204027}}.
	%Type = Article
	\bibitem[{Amelino-Camelia(2002)}]{Amelino-Camelia:2002aqz}
	\bibinfo{author}{G.~Amelino-Camelia},
	\newblock \bibinfo{title}{{Quantum gravity phenomenology: Status and
			prospects}},
	\newblock \bibinfo{journal}{Mod. Phys. Lett. A} \bibinfo{volume}{17}
	(\bibinfo{year}{2002}) \bibinfo{pages}{899--922}.
	\DOIprefix\doi{10.1142/S0217732302007612}.
	\href{http://arxiv.org/abs/gr-qc/0204051}{{\tt arXiv:gr-qc/0204051}}.
	%Type = Article
	\bibitem[{Ng(2003)}]{Ng:2003jk}
	\bibinfo{author}{Y.~J. Ng},
	\newblock \bibinfo{title}{{Selected topics in Planck scale physics}},
	\newblock \bibinfo{journal}{Mod. Phys. Lett. A} \bibinfo{volume}{18}
	(\bibinfo{year}{2003}) \bibinfo{pages}{1073--1098}.
	\DOIprefix\doi{10.1142/S0217732303010934}.
	\href{http://arxiv.org/abs/gr-qc/0305019}{{\tt arXiv:gr-qc/0305019}}.
	%Type = Article
	\bibitem[{Kostelecky and Mewes(2012)}]{Kostelecky:2011gq}
	\bibinfo{author}{A.~Kostelecky}, \bibinfo{author}{M.~Mewes},
	\newblock \bibinfo{title}{{Neutrinos with Lorentz-violating operators of
			arbitrary dimension}},
	\newblock \bibinfo{journal}{Phys. Rev. D} \bibinfo{volume}{85}
	(\bibinfo{year}{2012}) \bibinfo{pages}{096005}.
	\DOIprefix\doi{10.1103/PhysRevD.85.096005}.
	\href{http://arxiv.org/abs/1112.6395}{{\tt arXiv:1112.6395}}.
	%Type = Article
	\bibitem[{Barenboim et~al.(2019)Barenboim, Masud, Ternes, and
		T\'ortola}]{Barenboim:2018ctx}
	\bibinfo{author}{G.~Barenboim}, \bibinfo{author}{M.~Masud},
	\bibinfo{author}{C.~A. Ternes}, \bibinfo{author}{M.~T\'ortola},
	\newblock \bibinfo{title}{{Exploring the intrinsic Lorentz-violating parameters
			at DUNE}},
	\newblock \bibinfo{journal}{Phys. Lett. B} \bibinfo{volume}{788}
	(\bibinfo{year}{2019}) \bibinfo{pages}{308--315}.
	\DOIprefix\doi{10.1016/j.physletb.2018.11.040}.
	\href{http://arxiv.org/abs/1805.11094}{{\tt arXiv:1805.11094}}.
	%Type = Article
	\bibitem[{Kostelecky and Mewes(2004)}]{Kostelecky:2003cr}
	\bibinfo{author}{V.~A. Kostelecky}, \bibinfo{author}{M.~Mewes},
	\newblock \bibinfo{title}{{Lorentz and CPT violation in neutrinos}},
	\newblock \bibinfo{journal}{Phys. Rev. D} \bibinfo{volume}{69}
	(\bibinfo{year}{2004}) \bibinfo{pages}{016005}.
	\DOIprefix\doi{10.1103/PhysRevD.69.016005}.
	\href{http://arxiv.org/abs/hep-ph/0309025}{{\tt arXiv:hep-ph/0309025}}.
	%Type = Article
	\bibitem[{Greenberg(2002)}]{Greenberg:2002uu}
	\bibinfo{author}{O.~W. Greenberg},
	\newblock \bibinfo{title}{{CPT violation implies violation of Lorentz
			invariance}},
	\newblock \bibinfo{journal}{Phys. Rev. Lett.} \bibinfo{volume}{89}
	(\bibinfo{year}{2002}) \bibinfo{pages}{231602}.
	\DOIprefix\doi{10.1103/PhysRevLett.89.231602}.
	\href{http://arxiv.org/abs/hep-ph/0201258}{{\tt arXiv:hep-ph/0201258}}.
	%Type = Article
	\bibitem[{Appelquist and Carazzone(1975)}]{Appelquist:1974tg}
	\bibinfo{author}{T.~Appelquist}, \bibinfo{author}{J.~Carazzone},
	\newblock \bibinfo{title}{{Infrared Singularities and Massive Fields}},
	\newblock \bibinfo{journal}{Phys. Rev. D} \bibinfo{volume}{11}
	(\bibinfo{year}{1975}) \bibinfo{pages}{2856}.
	\DOIprefix\doi{10.1103/PhysRevD.11.2856}.
	%Type = Article
	\bibitem[{Colladay and Kostelecky(1998)}]{Colladay:1998fq}
	\bibinfo{author}{D.~Colladay}, \bibinfo{author}{V.~A. Kostelecky},
	\newblock \bibinfo{title}{{Lorentz violating extension of the standard model}},
	\newblock \bibinfo{journal}{Phys. Rev. D} \bibinfo{volume}{58}
	(\bibinfo{year}{1998}) \bibinfo{pages}{116002}.
	\DOIprefix\doi{10.1103/PhysRevD.58.116002}.
	\href{http://arxiv.org/abs/hep-ph/9809521}{{\tt arXiv:hep-ph/9809521}}.
	%Type = Article
	\bibitem[{Kostelecky and Lehnert(2001)}]{Kostelecky:2000mm}
	\bibinfo{author}{V.~A. Kostelecky}, \bibinfo{author}{R.~Lehnert},
	\newblock \bibinfo{title}{{Stability, causality, and Lorentz and CPT
			violation}},
	\newblock \bibinfo{journal}{Phys. Rev. D} \bibinfo{volume}{63}
	(\bibinfo{year}{2001}) \bibinfo{pages}{065008}.
	\DOIprefix\doi{10.1103/PhysRevD.63.065008}.
	\href{http://arxiv.org/abs/hep-th/0012060}{{\tt arXiv:hep-th/0012060}}.
	%Type = Article
	\bibitem[{Kostelecky(2004)}]{Kostelecky:2003fs}
	\bibinfo{author}{V.~A. Kostelecky},
	\newblock \bibinfo{title}{{Gravity, Lorentz violation, and the standard
			model}},
	\newblock \bibinfo{journal}{Phys. Rev. D} \bibinfo{volume}{69}
	(\bibinfo{year}{2004}) \bibinfo{pages}{105009}.
	\DOIprefix\doi{10.1103/PhysRevD.69.105009}.
	\href{http://arxiv.org/abs/hep-th/0312310}{{\tt arXiv:hep-th/0312310}}.
	%Type = Article
	\bibitem[{Bluhm(2006)}]{Bluhm:2005uj}
	\bibinfo{author}{R.~Bluhm},
	\newblock \bibinfo{title}{{Overview of the SME: Implications and phenomenology
			of Lorentz violation}},
	\newblock \bibinfo{journal}{Lect. Notes Phys.} \bibinfo{volume}{702}
	(\bibinfo{year}{2006}) \bibinfo{pages}{191--226}.
	\DOIprefix\doi{10.1007/3-540-34523-X_8}.
	\href{http://arxiv.org/abs/hep-ph/0506054}{{\tt arXiv:hep-ph/0506054}}.
	%Type = Article
	\bibitem[{Jacobson and Ohlsson(2004)}]{Jacobson:2003wc}
	\bibinfo{author}{M.~Jacobson}, \bibinfo{author}{T.~Ohlsson},
	\newblock \bibinfo{title}{{Extrinsic CPT violation in neutrino oscillations in
			matter}},
	\newblock \bibinfo{journal}{Phys. Rev. D} \bibinfo{volume}{69}
	(\bibinfo{year}{2004}) \bibinfo{pages}{013003}.
	\DOIprefix\doi{10.1103/PhysRevD.69.013003}.
	\href{http://arxiv.org/abs/hep-ph/0305064}{{\tt arXiv:hep-ph/0305064}}.
	%Type = Article
	\bibitem[{Ohlsson and Zhou(2015)}]{Ohlsson:2014cha}
	\bibinfo{author}{T.~Ohlsson}, \bibinfo{author}{S.~Zhou},
	\newblock \bibinfo{title}{{Extrinsic and Intrinsic CPT Asymmetries in Neutrino
			Oscillations}},
	\newblock \bibinfo{journal}{Nucl. Phys. B} \bibinfo{volume}{893}
	(\bibinfo{year}{2015}) \bibinfo{pages}{482--500}.
	\DOIprefix\doi{10.1016/j.nuclphysb.2015.02.015}.
	\href{http://arxiv.org/abs/1408.4722}{{\tt arXiv:1408.4722}}.
	%Type = Article
	\bibitem[{Kopp et~al.(2008)Kopp, Lindner, Ota, and Sato}]{Kopp:2007ne}
	\bibinfo{author}{J.~Kopp}, \bibinfo{author}{M.~Lindner},
	\bibinfo{author}{T.~Ota}, \bibinfo{author}{J.~Sato},
	\newblock \bibinfo{title}{{Non-standard neutrino interactions in reactor and
			superbeam experiments}},
	\newblock \bibinfo{journal}{Phys. Rev. D} \bibinfo{volume}{77}
	(\bibinfo{year}{2008}) \bibinfo{pages}{013007}.
	\DOIprefix\doi{10.1103/PhysRevD.77.013007}.
	\href{http://arxiv.org/abs/0708.0152}{{\tt arXiv:0708.0152}}.
	%Type = Article
	\bibitem[{Kumar et~al.(2021)Kumar, Khatun, Agarwalla, and
		Dighe}]{Kumar:2020wgz}
	\bibinfo{author}{A.~Kumar}, \bibinfo{author}{A.~Khatun}, \bibinfo{author}{S.~K.
		Agarwalla}, \bibinfo{author}{A.~Dighe},
	\newblock \bibinfo{title}{{From oscillation dip to oscillation valley in
			atmospheric neutrino experiments}},
	\newblock \bibinfo{journal}{Eur. Phys. J. C} \bibinfo{volume}{81}
	(\bibinfo{year}{2021}) \bibinfo{pages}{190}.
	\DOIprefix\doi{10.1140/epjc/s10052-021-08946-8}.
	\href{http://arxiv.org/abs/2006.14529}{{\tt arXiv:2006.14529}}.
	%Type = Misc
	\bibitem[{NuF(2020)}]{NuFIT}
	\bibinfo{title}{Nufit 5.0}, \bibinfo{year}{2020}. \URLprefix
	\url{http://www.nu-fit.org}.
	%Type = Article
	\bibitem[{Esteban et~al.(2020)Esteban, Gonzalez-Garcia, Maltoni, Schwetz, and
		Zhou}]{Esteban:2020cvm}
	\bibinfo{author}{I.~Esteban}, \bibinfo{author}{M.~C. Gonzalez-Garcia},
	\bibinfo{author}{M.~Maltoni}, \bibinfo{author}{T.~Schwetz},
	\bibinfo{author}{A.~Zhou},
	\newblock \bibinfo{title}{{The fate of hints: updated global analysis of
			three-flavor neutrino oscillations}},
	\newblock \bibinfo{journal}{JHEP} \bibinfo{volume}{09} (\bibinfo{year}{2020})
	\bibinfo{pages}{178}. \DOIprefix\doi{10.1007/JHEP09(2020)178}.
	\href{http://arxiv.org/abs/2007.14792}{{\tt arXiv:2007.14792}}.
	%Type = Article
	\bibitem[{de~Salas et~al.(2021)de~Salas, Forero, Gariazzo,
		Mart\'\i{}nez-Mirav\'e, Mena, Ternes, T\'ortola, and Valle}]{deSalas:2020pgw}
	\bibinfo{author}{P.~F. de~Salas}, \bibinfo{author}{D.~V. Forero},
	\bibinfo{author}{S.~Gariazzo}, \bibinfo{author}{P.~Mart\'\i{}nez-Mirav\'e},
	\bibinfo{author}{O.~Mena}, \bibinfo{author}{C.~A. Ternes},
	\bibinfo{author}{M.~T\'ortola}, \bibinfo{author}{J.~W.~F. Valle},
	\newblock \bibinfo{title}{{2020 global reassessment of the neutrino oscillation
			picture}},
	\newblock \bibinfo{journal}{JHEP} \bibinfo{volume}{02} (\bibinfo{year}{2021})
	\bibinfo{pages}{071}. \DOIprefix\doi{10.1007/JHEP02(2021)071}.
	\href{http://arxiv.org/abs/2006.11237}{{\tt arXiv:2006.11237}}.
	%Type = Article
	\bibitem[{Dziewonski and Anderson(1981)}]{Dziewonski:1981xy}
	\bibinfo{author}{A.~Dziewonski}, \bibinfo{author}{D.~Anderson},
	\newblock \bibinfo{title}{{Preliminary Reference Earth Model}},
	\newblock \bibinfo{journal}{Phys.Earth Planet.Interiors} \bibinfo{volume}{25}
	(\bibinfo{year}{1981}) \bibinfo{pages}{297--356}.
	\DOIprefix\doi{10.1016/0031-9201(81)90046-7}.
	%Type = Article
	\bibitem[{Weinberg(1979)}]{Weinberg:1979sa}
	\bibinfo{author}{S.~Weinberg},
	\newblock \bibinfo{title}{{Baryon and Lepton Nonconserving Processes}},
	\newblock \bibinfo{journal}{Phys. Rev. Lett.} \bibinfo{volume}{43}
	(\bibinfo{year}{1979}) \bibinfo{pages}{1566--1570}.
	\DOIprefix\doi{10.1103/PhysRevLett.43.1566}.
	%Type = Article
	\bibitem[{Abbasi et~al.(2022)}]{IceCube:2022ubv}
	\bibinfo{author}{R.~Abbasi}, et~al. (\bibinfo{collaboration}{IceCube}),
	\newblock \bibinfo{title}{{Strong Constraints on Neutrino Nonstandard
			Interactions from TeV-Scale $\nu_u$ Disappearance at IceCube}},
	\newblock \bibinfo{journal}{Phys. Rev. Lett.} \bibinfo{volume}{129}
	(\bibinfo{year}{2022}) \bibinfo{pages}{011804}.
	\DOIprefix\doi{10.1103/PhysRevLett.129.011804}.
	\href{http://arxiv.org/abs/2201.03566}{{\tt arXiv:2201.03566}}.
	%Type = Article
	\bibitem[{Diaz(2015)}]{Diaz:2015dxa}
	\bibinfo{author}{J.~S. Diaz},
	\newblock \bibinfo{title}{{Correspondence between nonstandard interactions and
			CPT violation in neutrino oscillations}}  (\bibinfo{year}{2015}).
	\href{http://arxiv.org/abs/1506.01936}{{\tt arXiv:1506.01936}}.
	%Type = Article
	\bibitem[{Agarwalla and Masud(2020)}]{KumarAgarwalla:2019gdj}
	\bibinfo{author}{S.~K. Agarwalla}, \bibinfo{author}{M.~Masud},
	\newblock \bibinfo{title}{{Can Lorentz invariance violation affect the
			sensitivity of deep underground neutrino experiment?}},
	\newblock \bibinfo{journal}{Eur. Phys. J. C} \bibinfo{volume}{80}
	(\bibinfo{year}{2020}) \bibinfo{pages}{716}.
	\DOIprefix\doi{10.1140/epjc/s10052-020-8303-1}.
	\href{http://arxiv.org/abs/1912.13306}{{\tt arXiv:1912.13306}}.
	%Type = Article
	\bibitem[{Gandhi et~al.(2006)Gandhi, Ghoshal, Goswami, Mehta, and
		Sankar}]{Gandhi:2004bj}
	\bibinfo{author}{R.~Gandhi}, \bibinfo{author}{P.~Ghoshal},
	\bibinfo{author}{S.~Goswami}, \bibinfo{author}{P.~Mehta},
	\bibinfo{author}{S.~U. Sankar},
	\newblock \bibinfo{title}{{Earth matter effects at very long baselines and the
			neutrino mass hierarchy}},
	\newblock \bibinfo{journal}{Phys. Rev. D} \bibinfo{volume}{73}
	(\bibinfo{year}{2006}) \bibinfo{pages}{053001}.
	\DOIprefix\doi{10.1103/PhysRevD.73.053001}.
	\href{http://arxiv.org/abs/hep-ph/0411252}{{\tt arXiv:hep-ph/0411252}}.
	%Type = Phdthesis
	\bibitem[{Agarwalla(2008)}]{Agarwalla:2008jin}
	\bibinfo{author}{S.~K. Agarwalla}, \bibinfo{title}{{Some Aspects of Neutrino
			Mixing and Oscillations}}, Ph.D. thesis, Calcutta U., \bibinfo{year}{2008}.
	\href{http://arxiv.org/abs/0908.4267}{{\tt arXiv:0908.4267}}.
	%Type = Article
	\bibitem[{Roe(2017)}]{Roe:2017zdw}
	\bibinfo{author}{B.~Roe},
	\newblock \bibinfo{title}{{Matter density versus distance for the neutrino beam
			from Fermilab to Lead, South Dakota, and comparison of oscillations with
			variable and constant density}},
	\newblock \bibinfo{journal}{Phys. Rev. D} \bibinfo{volume}{95}
	(\bibinfo{year}{2017}) \bibinfo{pages}{113004}.
	\DOIprefix\doi{10.1103/PhysRevD.95.113004}.
	\href{http://arxiv.org/abs/1707.02322}{{\tt arXiv:1707.02322}}.
	%Type = Article
	\bibitem[{Abi et~al.(2021)}]{DUNE:2021cuw}
	\bibinfo{author}{B.~Abi}, et~al. (\bibinfo{collaboration}{DUNE}),
	\newblock \bibinfo{title}{{Experiment Simulation Configurations Approximating
			DUNE TDR}}  (\bibinfo{year}{2021}).
	\href{http://arxiv.org/abs/2103.04797}{{\tt arXiv:2103.04797}}.
	%Type = Article
	\bibitem[{Mikheev and Smirnov(1985)}]{Mikheev:1986gs}
	\bibinfo{author}{S.~P. Mikheev}, \bibinfo{author}{A.~Y. Smirnov},
	\newblock \bibinfo{title}{{Resonance enhancement of oscillations in matter and
			solar neutrino spectroscopy}},
	\newblock \bibinfo{journal}{Sov. J. Nucl. Phys.} \bibinfo{volume}{42}
	(\bibinfo{year}{1985}) \bibinfo{pages}{913}.
	\bibinfo{note}{[Yad.Fiz.42:1441-1448,1985]}.
	%Type = Article
	\bibitem[{Mikheev and Smirnov(1986)}]{Mikheev:1986wj}
	\bibinfo{author}{S.~Mikheev}, \bibinfo{author}{A.~Y. Smirnov},
	\newblock \bibinfo{title}{{Resonant amplification of neutrino oscillations in
			matter and solar neutrino spectroscopy}},
	\newblock \bibinfo{journal}{Nuovo Cim.} \bibinfo{volume}{C9}
	(\bibinfo{year}{1986}) \bibinfo{pages}{17}.
	\DOIprefix\doi{10.1007/BF02508049}.
	%Type = Article
	\bibitem[{Petcov(1998)}]{Petcov:1998su}
	\bibinfo{author}{S.~Petcov},
	\newblock \bibinfo{title}{{Diffractive - like (or parametric resonance - like?)
			enhancement of the earth (day - night) effect for solar neutrinos crossing
			the earth core}},
	\newblock \bibinfo{journal}{Phys. Lett. B} \bibinfo{volume}{434}
	(\bibinfo{year}{1998}) \bibinfo{pages}{321}.
	\DOIprefix\doi{10.1016/S0370-2693(98)00742-4}.
	\href{http://arxiv.org/abs/hep-ph/9805262}{{\tt arXiv:hep-ph/9805262}}.
	%Type = Article
	\bibitem[{Chizhov et~al.(1998)Chizhov, Maris, and Petcov}]{Chizhov:1998ug}
	\bibinfo{author}{M.~Chizhov}, \bibinfo{author}{M.~Maris},
	\bibinfo{author}{S.~T. Petcov},
	\newblock \bibinfo{title}{{On the oscillation length resonance in the
			transitions of solar and atmospheric neutrinos crossing the earth core}}
	(\bibinfo{year}{1998}). \href{http://arxiv.org/abs/hep-ph/9810501}{{\tt
			arXiv:hep-ph/9810501}}.
	%Type = Article
	\bibitem[{Petcov(1999)}]{Petcov:1998sg}
	\bibinfo{author}{S.~T. Petcov},
	\newblock \bibinfo{title}{{New enhancement mechanism of the transitions in the
			earth of the solar and atmospheric neutrinos crossing the earth core}},
	\newblock \bibinfo{journal}{Nucl. Phys. B Proc. Suppl.} \bibinfo{volume}{77}
	(\bibinfo{year}{1999}) \bibinfo{pages}{93--97}.
	\DOIprefix\doi{10.1016/S0920-5632(99)00403-X}.
	\href{http://arxiv.org/abs/hep-ph/9809587}{{\tt arXiv:hep-ph/9809587}}.
	%Type = Article
	\bibitem[{Chizhov and Petcov(1999)}]{Chizhov:1999az}
	\bibinfo{author}{M.~Chizhov}, \bibinfo{author}{S.~Petcov},
	\newblock \bibinfo{title}{{New conditions for a total neutrino conversion in a
			medium}},
	\newblock \bibinfo{journal}{Phys.Rev.Lett.} \bibinfo{volume}{83}
	(\bibinfo{year}{1999}) \bibinfo{pages}{1096--1099}.
	\DOIprefix\doi{10.1103/PhysRevLett.83.1096}.
	\href{http://arxiv.org/abs/hep-ph/9903399}{{\tt arXiv:hep-ph/9903399}}.
	%Type = Article
	\bibitem[{Chizhov and Petcov(2001)}]{Chizhov:1999he}
	\bibinfo{author}{M.~V. Chizhov}, \bibinfo{author}{S.~T. Petcov},
	\newblock \bibinfo{title}{{Enhancing mechanisms of neutrino transitions in a
			medium of nonperiodic constant density layers and in the earth}},
	\newblock \bibinfo{journal}{Phys. Rev. D} \bibinfo{volume}{63}
	(\bibinfo{year}{2001}) \bibinfo{pages}{073003}.
	\DOIprefix\doi{10.1103/PhysRevD.63.073003}.
	\href{http://arxiv.org/abs/hep-ph/9903424}{{\tt arXiv:hep-ph/9903424}}.
	%Type = Article
	\bibitem[{Akhmedov(1999)}]{Akhmedov:1998ui}
	\bibinfo{author}{E.~K. Akhmedov},
	\newblock \bibinfo{title}{{Parametric resonance of neutrino oscillations and
			passage of solar and atmospheric neutrinos through the earth}},
	\newblock \bibinfo{journal}{Nucl. Phys.} \bibinfo{volume}{B538}
	(\bibinfo{year}{1999}) \bibinfo{pages}{25}.
	\DOIprefix\doi{10.1016/S0550-3213(98)00723-8}.
	\href{http://arxiv.org/abs/hep-ph/9805272}{{\tt arXiv:hep-ph/9805272}}.
	%Type = Article
	\bibitem[{Akhmedov et~al.(1999)Akhmedov, Dighe, Lipari, and
		Smirnov}]{Akhmedov:1998xq}
	\bibinfo{author}{E.~K. Akhmedov}, \bibinfo{author}{A.~Dighe},
	\bibinfo{author}{P.~Lipari}, \bibinfo{author}{A.~Smirnov},
	\newblock \bibinfo{title}{{Atmospheric neutrinos at Super-Kamiokande and
			parametric resonance in neutrino oscillations}},
	\newblock \bibinfo{journal}{Nucl. Phys.} \bibinfo{volume}{B542}
	(\bibinfo{year}{1999}) \bibinfo{pages}{3}.
	\DOIprefix\doi{10.1016/S0550-3213(98)00825-6}.
	\href{http://arxiv.org/abs/hep-ph/9808270}{{\tt arXiv:hep-ph/9808270}}.
	%Type = Article
	\bibitem[{Chatterjee et~al.(2014)Chatterjee, Meghna, Rawat, Thakore, Bhatnagar
		et~al.}]{Chatterjee:2014vta}
	\bibinfo{author}{A.~Chatterjee}, \bibinfo{author}{K.~Meghna},
	\bibinfo{author}{K.~Rawat}, \bibinfo{author}{T.~Thakore},
	\bibinfo{author}{V.~Bhatnagar}, et~al.,
	\newblock \bibinfo{title}{{A Simulations Study of the Muon Response of the Iron
			Calorimeter Detector at the India-based Neutrino Observatory}},
	\newblock \bibinfo{journal}{JINST} \bibinfo{volume}{9} (\bibinfo{year}{2014})
	\bibinfo{pages}{P07001}. \DOIprefix\doi{10.1088/1748-0221/9/07/P07001}.
	\href{http://arxiv.org/abs/1405.7243}{{\tt arXiv:1405.7243}}.
	%Type = Article
	\bibitem[{Devi et~al.(2013)Devi, Ghosh, Kaur, Mohan, Choubey
		et~al.}]{Devi:2013wxa}
	\bibinfo{author}{M.~M. Devi}, \bibinfo{author}{A.~Ghosh},
	\bibinfo{author}{D.~Kaur}, \bibinfo{author}{L.~S. Mohan},
	\bibinfo{author}{S.~Choubey}, et~al.,
	\newblock \bibinfo{title}{{Hadron energy response of the Iron Calorimeter
			detector at the India-based Neutrino Observatory}},
	\newblock \bibinfo{journal}{JINST} \bibinfo{volume}{8} (\bibinfo{year}{2013})
	\bibinfo{pages}{P11003}. \DOIprefix\doi{10.1088/1748-0221/8/11/P11003}.
	\href{http://arxiv.org/abs/1304.5115}{{\tt arXiv:1304.5115}}.
	%Type = Article
	\bibitem[{Casper(2002)}]{Casper:2002sd}
	\bibinfo{author}{D.~Casper},
	\newblock \bibinfo{title}{{The Nuance neutrino physics simulation, and the
			future}},
	\newblock \bibinfo{journal}{Nucl. Phys. Proc. Suppl.} \bibinfo{volume}{112}
	(\bibinfo{year}{2002}) \bibinfo{pages}{161}.
	\DOIprefix\doi{10.1016/S0920-5632(02)01756-5}.
	\href{http://arxiv.org/abs/hep-ph/0208030}{{\tt arXiv:hep-ph/0208030}}.
	%Type = Article
	\bibitem[{Honda et~al.(2015)Honda, Sajjad~Athar, Kajita, Kasahara, and
		Midorikawa}]{Honda:2015fha}
	\bibinfo{author}{M.~Honda}, \bibinfo{author}{M.~Sajjad~Athar},
	\bibinfo{author}{T.~Kajita}, \bibinfo{author}{K.~Kasahara},
	\bibinfo{author}{S.~Midorikawa},
	\newblock \bibinfo{title}{{Atmospheric neutrino flux calculation using the
			NRLMSISE-00 atmospheric model}},
	\newblock \bibinfo{journal}{Phys. Rev. D} \bibinfo{volume}{92}
	(\bibinfo{year}{2015}) \bibinfo{pages}{023004}.
	\DOIprefix\doi{10.1103/PhysRevD.92.023004}.
	\href{http://arxiv.org/abs/1502.03916}{{\tt arXiv:1502.03916}}.
	%Type = Article
	\bibitem[{Ghosh et~al.(2013)Ghosh, Thakore, and Choubey}]{Ghosh:2012px}
	\bibinfo{author}{A.~Ghosh}, \bibinfo{author}{T.~Thakore},
	\bibinfo{author}{S.~Choubey},
	\newblock \bibinfo{title}{{Determining the Neutrino Mass Hierarchy with INO,
			T2K, NOvA and Reactor Experiments}},
	\newblock \bibinfo{journal}{JHEP} \bibinfo{volume}{1304} (\bibinfo{year}{2013})
	\bibinfo{pages}{009}. \DOIprefix\doi{10.1007/JHEP04(2013)009}.
	\href{http://arxiv.org/abs/1212.1305}{{\tt arXiv:1212.1305}}.
	%Type = Article
	\bibitem[{Thakore et~al.(2013)Thakore, Ghosh, Choubey, and
		Dighe}]{Thakore:2013xqa}
	\bibinfo{author}{T.~Thakore}, \bibinfo{author}{A.~Ghosh},
	\bibinfo{author}{S.~Choubey}, \bibinfo{author}{A.~Dighe},
	\newblock \bibinfo{title}{{The Reach of INO for Atmospheric Neutrino
			Oscillation Parameters}},
	\newblock \bibinfo{journal}{JHEP} \bibinfo{volume}{1305} (\bibinfo{year}{2013})
	\bibinfo{pages}{058}. \DOIprefix\doi{10.1007/JHEP05(2013)058}.
	\href{http://arxiv.org/abs/1303.2534}{{\tt arXiv:1303.2534}}.
	%Type = Article
	\bibitem[{Devi et~al.(2014)Devi, Thakore, Agarwalla, and Dighe}]{Devi:2014yaa}
	\bibinfo{author}{M.~M. Devi}, \bibinfo{author}{T.~Thakore},
	\bibinfo{author}{S.~K. Agarwalla}, \bibinfo{author}{A.~Dighe},
	\newblock \bibinfo{title}{{Enhancing sensitivity to neutrino parameters at INO
			combining muon and hadron information}},
	\newblock \bibinfo{journal}{JHEP} \bibinfo{volume}{10} (\bibinfo{year}{2014})
	\bibinfo{pages}{189}. \DOIprefix\doi{10.1007/JHEP10(2014)189}.
	\href{http://arxiv.org/abs/1406.3689}{{\tt arXiv:1406.3689}}.
	%Type = Article
	\bibitem[{Upadhyay et~al.(2022)Upadhyay, Kumar, Agarwalla, and
		Dighe}]{Upadhyay:2022jfd}
	\bibinfo{author}{A.~K. Upadhyay}, \bibinfo{author}{A.~Kumar},
	\bibinfo{author}{S.~K. Agarwalla}, \bibinfo{author}{A.~Dighe},
	\newblock \bibinfo{title}{{Locating the Core-Mantle Boundary using Oscillations
			of Atmospheric Neutrinos}}  (\bibinfo{year}{2022}).
	\href{http://arxiv.org/abs/2211.08688}{{\tt arXiv:2211.08688}}.
	%Type = Article
	\bibitem[{Baker and Cousins(1984)}]{Baker:1983tu}
	\bibinfo{author}{S.~Baker}, \bibinfo{author}{R.~D. Cousins},
	\newblock \bibinfo{title}{{Clarification of the Use of Chi Square and
			Likelihood Functions in Fits to Histograms}},
	\newblock \bibinfo{journal}{Nucl. Instrum. Meth.} \bibinfo{volume}{221}
	(\bibinfo{year}{1984}) \bibinfo{pages}{437--442}.
	\DOIprefix\doi{10.1016/0167-5087(84)90016-4}.
	%Type = Article
	\bibitem[{Blennow et~al.(2014)Blennow, Coloma, Huber, and
		Schwetz}]{Blennow:2013oma}
	\bibinfo{author}{M.~Blennow}, \bibinfo{author}{P.~Coloma},
	\bibinfo{author}{P.~Huber}, \bibinfo{author}{T.~Schwetz},
	\newblock \bibinfo{title}{{Quantifying the sensitivity of oscillation
			experiments to the neutrino mass ordering}},
	\newblock \bibinfo{journal}{JHEP} \bibinfo{volume}{1403} (\bibinfo{year}{2014})
	\bibinfo{pages}{028}. \DOIprefix\doi{10.1007/JHEP03(2014)028}.
	\href{http://arxiv.org/abs/1311.1822}{{\tt arXiv:1311.1822}}.
	%Type = Phdthesis
	\bibitem[{Kameda(2002)}]{Kameda:2002fx}
	\bibinfo{author}{J.~Kameda}, \bibinfo{title}{{Detailed studies of neutrino
			oscillations with atmospheric neutrinos of wide energy range from 100 MeV to
			1000 GeV in Super-Kamiokande}}, Ph.D. thesis, Tokyo U., \bibinfo{year}{2002}.
	%Type = Article
	\bibitem[{Gonzalez-Garcia and
		Maltoni(2004{\natexlab{a}})}]{Gonzalez-Garcia:2004pka}
	\bibinfo{author}{M.~C. Gonzalez-Garcia}, \bibinfo{author}{M.~Maltoni},
	\newblock \bibinfo{title}{{Atmospheric neutrino oscillations and new physics}},
	\newblock \bibinfo{journal}{Phys. Rev. D} \bibinfo{volume}{70}
	(\bibinfo{year}{2004}{\natexlab{a}}) \bibinfo{pages}{033010}.
	\DOIprefix\doi{10.1103/PhysRevD.70.033010}.
	\href{http://arxiv.org/abs/hep-ph/0404085}{{\tt arXiv:hep-ph/0404085}}.
	%Type = Article
	\bibitem[{Gonzalez-Garcia and
		Maltoni(2004{\natexlab{b}})}]{GonzalezGarcia:2004wg}
	\bibinfo{author}{M.~C. Gonzalez-Garcia}, \bibinfo{author}{M.~Maltoni},
	\newblock \bibinfo{title}{{Atmospheric neutrino oscillations and new physics}},
	\newblock \bibinfo{journal}{Phys. Rev.} \bibinfo{volume}{D70}
	(\bibinfo{year}{2004}{\natexlab{b}}) \bibinfo{pages}{033010}.
	\DOIprefix\doi{10.1103/PhysRevD.70.033010}.
	\href{http://arxiv.org/abs/hep-ph/0404085}{{\tt arXiv:hep-ph/0404085}}.
	%Type = Article
	\bibitem[{Huber et~al.(2002)Huber, Lindner, and Winter}]{Huber:2002mx}
	\bibinfo{author}{P.~Huber}, \bibinfo{author}{M.~Lindner},
	\bibinfo{author}{W.~Winter},
	\newblock \bibinfo{title}{{Superbeams versus neutrino factories}},
	\newblock \bibinfo{journal}{Nucl. Phys.} \bibinfo{volume}{B645}
	(\bibinfo{year}{2002}) \bibinfo{pages}{3--48}.
	\DOIprefix\doi{10.1016/S0550-3213(02)00825-8}.
	\href{http://arxiv.org/abs/hep-ph/0204352}{{\tt arXiv:hep-ph/0204352}}.
	%Type = Article
	\bibitem[{Fogli et~al.(2002)Fogli, Lisi, Marrone, Montanino, and
		Palazzo}]{Fogli:2002pt}
	\bibinfo{author}{G.~L. Fogli}, \bibinfo{author}{E.~Lisi},
	\bibinfo{author}{A.~Marrone}, \bibinfo{author}{D.~Montanino},
	\bibinfo{author}{A.~Palazzo},
	\newblock \bibinfo{title}{{Getting the most from the statistical analysis of
			solar neutrino oscillations}},
	\newblock \bibinfo{journal}{Phys. Rev. D} \bibinfo{volume}{66}
	(\bibinfo{year}{2002}) \bibinfo{pages}{053010}.
	\DOIprefix\doi{10.1103/PhysRevD.66.053010}.
	\href{http://arxiv.org/abs/hep-ph/0206162}{{\tt arXiv:hep-ph/0206162}}.
\end{thebibliography}

%=============================%

\end{document}